%% file: main.tex
\documentclass[aps,prd,superscriptaddress,showkeys,showpacs,twocolumn,longbibliography]{revtex4-2}
\usepackage{color}
\usepackage{graphicx}
\usepackage{subcaption}
\usepackage{booktabs}
\usepackage{multirow}
\usepackage{appendix}
\usepackage{amsmath}
\usepackage{comment}
\usepackage{float}
\usepackage{inputenc}
\usepackage{fontenc}
\usepackage[bookmarks,bookmarksnumbered]{hyperref}
\hypersetup{colorlinks = true,
            linkcolor = blue,
            anchorcolor =red,
            citecolor = blue,
            pdfauthor=author}

\usepackage[dvipsnames]{xcolor}

\graphicspath{{figs/}}

\newcommand{\jetscape}{\textsc{jetscape}}
\newcommand{\martini}{\textsc{martini}}
\newcommand{\music}{\textsc{music}}
\newcommand{\cujet}{\textsc{cujet}}
\newcommand{\matter}{\textsc{matter}}
\newcommand{\pythia}{\textsc{pythia}}

\newcommand{\amy}{\textsc{amy}}
\newcommand{\dglv}{\textsc{dglv}}
\newcommand{\trento}{\textsc{trento}}
\newcommand{\ipg}{\textsc{ip-glasma}}
\newcommand{\vishnu}{\textsc{vishnu}}
\newcommand{\alphas}{\ensuremath{\alpha_\mathrm{s}}}
\newcommand{\raa}{\ensuremath{R_{AA}}}
\newcommand{\lbt}{\textsc{lbt}}

\newcommand{\pbpb}{\ensuremath{\mathrm{Pb+Pb}}}
\newcommand{\pp}{\ensuremath{\mathrm{p+p}}}

\begin{document}
\title{QCD jets in a hot and dense medium: a study of shower formation time and collision kernels}

\author{Rouzbeh Modarresi-Yazdi}
\email{rouzbeh.modarresi-yazdi@mail.mcgill.ca}
\affiliation{Department of Physics, McGill University, 3600 University street, Montreal, QC, Canada H3A 2T8}
\author{Shuzhe Shi}
\email[]{shuzhe-shi@tsinghua.edu.cn}
\affiliation{Department of Physics, Tsinghua University, Beijing 100084, China}
\affiliation{Center for Nuclear Theory, Department of Physics and Astronomy, Stony Brook University, Stony Brook, New York 11794-3800, USA}
\author{Charles Gale}
\email{charles.gale@mcgill.ca}
\affiliation{Department of Physics, McGill University, 3600 University street, Montreal, QC, Canada H3A 2T8}
\author{Sangyong Jeon}
\email{sangyong.jeon@mcgill.ca}
\affiliation{Department of Physics, McGill University, 3600 University street, Montreal, QC, Canada H3A 2T8}

\begin{abstract} 

With the use of \martini, a model which considers evolving QCD jets against a fluid dynamical background, it is shown that the introduction of formation time to the parton shower after the initial hard scattering is essential for a simultaneous description of charged hadron and jet $R_{AA}$. This inclusion also improves jet shape ratio at small angle and jet fragmentation function ratios of leading charged hadrons. The \martini\, framework is then aimed at a study of the leading order, next-to-leading-order, and non perturbative collision kernels. Sizable differences in the modification of jet substructure observables, i.e., jet shape and fragmentation functions are observed. Such differences are caused by the difference in the radiation rates of relatively soft gluons that survive in the evolution in medium.

\end{abstract}

\maketitle
        \input{introduction}
        \input{eloss_theory}
        \input{evolving_sim}
        
    \section{Results}
        This section contains the results of our simulations and associated discussions. This is done in two steps: Sec.~\ref{sec:fit.and.formtime} discusses the results of the inclusion of formation-time in the high virtuality stage of the parton shower, with focus on charged hadron and jet observables. Sec.~\ref{sec:higher.order.kernels} analyzes the effects of using the higher order collision kernels in calculations of hard parton quenching, with high-virtuality parton showers that include formation-time. The system under consideration throughout this work is \pbpb\, collisions at $2.76$~ATeV. 
            
        \input{section_formation_time}

        \input{section_high_order_kernels}

        \input{conclusion}

\acknowledgments
    The authors acknowledge useful discussions with Shanshan Cao, Matthew Heffernan, Weiyao Ke, Amit Kumar, and Abhijit Majumder. This work is supported in part by the Natural Sciences and Engineering Research Council of Canada (RMY, CG, SJ), in part by Tsinghua University under grant No. 53330500923 (SS), and in part by the U.S. Department of Energy, Office of Science, Office of Nuclear Physics, under grant No. DE-FG88ER41450 (SS). Computations were carried out on the Narval, Beluga and Graham clusters managed by Calcul Qu\'ebec and the Digital Research Alliance of Canada. 
    \bibliographystyle{apsrev4-1.bst}
    \bibliography{biblio}
    \begin{appendix}
        \input{appendix_comp_matter}
        \input{appendix_coll_kern_long_etime}
        \input{appendix_pythia_guns}
    \end{appendix}
    
\end{document}

%% file: introduction.tex
\section{Introduction}\label{sec:introduction}
The production and characterization of hot and dense strongly interacting matter is currently a mainstream area of research in subatomic physics. One of the goals of this research is to establish a faithful representation of the QCD phase diagram, of which little is currently known. The presence of exotic phase, the quark-gluon plasma (QGP) was a prediction of non-perturbative QCD calculations~\cite{[{See, for example, }][{, and references therein.}]Ratti:2018ksb}, and its existence has been confirmed by experiments done at the Relativistic Heavy-Ion Collider (RHIC) and later reinforced by results obtained at the Large Hadron Collider (LHC)~\cite{[{See, for example, }][{, and references therein.}]Busza:2018rrf}. The experimental results were mostly obtained by colliding different nuclear species at relativistic energies, but more recently QGP characteristic signatures were also observed in collisions involving light nuclei and even protons \cite{Jacazio:2024qpb}. 

It is fair to write that the physics of the QGP has now entered an era of characterization through precision studies often involving differential observables and penetrating probes. For instance, the ``temperature'' of the plasma can be assessed by measuring electromagnetic radiation emitted throughout the space-time evolution of the colliding system. This radiation can take the form of real photons~\cite{Shen:2013vja} or of lepton pairs~\cite{Churchill:2023zkk}. Another class of probes that can provide tomographic information is that associated with QCD jets~\cite{Connors:2017ptx}: the spray of particles associated with the decay of energetic off-shell partons formed in the very early collisions. The jets are reconstituted by measuring correlated hadrons in the final state, and they can provide unique information about the medium with which they interact. Jets and medium can exchange energy, and the medium can alter and distort the original jet shape. 

Because the QCD medium created in heavy-ion collisions is rapidly evolving through several stages -- some of which may be further away from equilibrium (thermal and/or chemical) than others -- it has rapidly become clear that theoretical analyses needed to include some realistic modeling of the time evolution. This includes approaches which simulate how the medium as a whole evolve in space and in time, and how the partons and the background matter exchange energy and momentum. 

There are now several theoretical models which evolve jets against a fluid dynamical background. A few examples include \martini~\cite{Schenke:2009gb}, \lbt\,~(Linearized Boltzmann Transport)~\cite{Li:2010ts, Guo:2000nz, Wang:2001ifa, Zhang:2003yn, Schafer:2007xh, He:2015pra, Cao:2016gvr}, \cujet~\cite{Buzzatti:2011vt, Xu:2014ica, Shi:2018izg, Shi:2018lsf}, and \textsc{AdS-CFT}~\cite{Casalderrey-Solana:2014bpa}. Each of those approaches depends on the details of the theoretical treatment of jet energy-loss. Some formalisms that aim to tackle this are \amy~\cite{Arnold:2002ja, Jeon:2003gi}, higher twist~\cite{Qiu:1990xxa, Qiu:1990xy, Majumder:2010qh}, and \dglv~\cite{Gyulassy:1993hr,Wang:1994fx,Gyulassy:2000er, Djordjevic:2003zk}. For a recent review of the various frameworks and models of jet energy-loss, see Ref.~\cite{Cao:2024pxc} and references therein.

Having noted the importance of the precision modeling of relativistic heavy-ion collisions, we focus in this work on two somewhat technical but nevertheless important aspects. We choose to use \martini, together with the parton emission rates associated with \amy. The first topic of investigation has to do with the early-time dynamics of jet evolution. In the chronology of a heavy-ion collision, the very first hard interactions will generate a parton shower~\cite{[{See, for example, }][{, and references therein.}]Ellis:1996mzs}. Depending on the energy of the scattering and the timescale associated with the development of the medium, this parton shower may, or may not, have time to evolve into an ensemble of on-shell partons before it finds itself embedded into the strongly interacting fluid. Thus there are two broad classes of models that can be considered for jet energy loss. In the first, the high virtuality parton shower is assumed to have fully developed before the onset of hydrodynamics. One can then make the approximation that the shower was fully developed at time $\tau=0^+$ and simply allow the hard partons to free stream until $\tau = \tau_{\mathrm{hydro}}$, the start time of the hydrodynamic evolution. In effect, the assumption of the model is that of a ``vacuum-like'' parton shower with an instantaneous or near-instantaneous development. In the second class of models, the final state parton shower post hard scattering evolves in time. Thus some hard partons shed their virtuality before the start time of hydrodynamics, some during and some may even come on-shell after the QGP has evaporated. We can then consider such models as those with a ``time-delayed'' parton shower.   

The first of the above is known as a ``single-stage'' simulation of jet energy loss and is how several energy loss models were previously used. In this type of model, it is assumed that the energy loss of the jet during the high virtuality phase of its evolution is negligible, and that the parton shower would develop rapidly, mostly before the onset of hydrodynamics and QGP evolution. 
In the second case, the high-virtuality final state shower evolution is allowed to develop as a function of time, and some parts are still forming as the hadronic fluid appears and evolves. We shall see that this element is crucial in the interpretation of several jet-related observables. 

After the study and discussion of formation time in the high-virtuality stage, we turn our attention to the effect of the new, higher-order collision kernels on the \amy\, radiative rates. Studies of these kernels have so far seen them used in computations of the leading-order \amy\, rates and then compared in simulations of parton energy loss in a QGP brick~\cite{Schlichting:2021idr} and in simulations with a dynamically evolving QGP~\cite{Yazdi:2022bru}. In the latter, the strong coupling used in the jet energy loss channels was held fixed so as to simplify the comparison of the different kernels. Here, this restriction is relaxed and scale dependence is introduced to \alphas. 

This paper is organized as follows: the next section introduces briefly the \martini\, framework, and the formalism used to treat parton energy loss. We then discuss further details of how partons are evolved in \martini\ and how parton formation time is applied to the partonic shower development. The section on results features evaluations of the nuclear modification factor for charged particles and reconstructed jets, as well as calculations of jet substructure ratios. We also highlight differences and similarities between results obtained with different scattering kernels. We finish with a conclusion and an outlook. 

%% file: eloss_theory.tex
\section{Jet energy loss in MARTINI}\label{sec:eloss.theory}  
    The \martini\, framework~\cite{Schenke:2009gb} is the Monte Carlo solution to the rate equation for the evolving hard parton distributions, representing energy gain and loss
        \begin{equation}
            \begin{split}
                \frac{df}{dt}(p) = \int_{-\infty}^{\infty} dk\Big[ &\frac{d\Gamma(p+k,k)}{dk} f(p+k)\\
                 & - \frac{d\Gamma(p,k)}{dk} f(p)\Big],
            \end{split}
        \end{equation}
    where $f(p)$ is the distribution of the evolving hard partons, quarks and gluons, in the medium and $d\Gamma/dk$ are the energy loss rates, radiative or elastic. The framework includes the two dominant modes of energy loss available to a hard parton in a QGP medium: gluon bremsstrahlung (or inelastic splittings) and elastic collisions with the medium particles. \martini\, treats both these energy loss channels simultaneously. In the following sections, we discuss the different energy loss channels used in \martini.

    \subsection{Radiative Energy Loss}\label{sec:rad.energy.loss.theory}
        The main channel of energy loss is the inelastic branching or the \textit{radiative} channel. \martini\, uses the rates computed using the \amy\; formalism~\cite{Arnold:2001ba,Arnold:2001ms,Arnold:2002ja,Jeon:2003gi}. These rates are derived in the limit of an infinite QGP at local thermal equilibrium. As such the rates are not time dependent and can be thought of as being to all orders in opacity. It is further assumed that the QGP temperature is asymptotically high and thus the strong coupling, $g$, is a small parameter. The inelastic rates are given by~\cite{Schenke:2009gb, Yazdi:2022bru}
        \begin{equation}
        \begin{split}
            \frac{\mathrm{d} \Gamma_{i\to jk}}{\mathrm{d} x} (p,x) =\;& 
            	\frac{\alpha_s P_{i \to jk}(x)}{[2p\,x(1{-}x)]^2} \bar{f}_j(x\,p)\, \bar{f}_k((1-x)p)\nonumber\\
             &\; \times\int \! \frac{\mathrm{d}^2 \mathbf{h}_{\perp}}{(2\pi)^2} ~\text{Re} \left[ 2\mathbf{h}_{\perp} \cdot \mathbf{g}_{(x,p)}(\mathbf{h}_{\perp}) \right].
             \label{eq:splitting.rate.martini}
        \end{split}
        \end{equation}
        In the above, $\bar{f}_{j} = (1\pm f_{j})$ denotes the distribution function of parton $j$, including the Bose enhancement or Pauli suppression factor. $P_{i\to jk}(x)$ represents the DGLAP splitting kernel for an incoming hard parton $i$ of momentum $p$ to split to parton $j$, carrying momentum $xp$ and parton $k$ with momentum $(1-x)p$. The function $\mathbf{g}_{(x,p)}(\mathbf{h}_{\perp})$ encodes information about the transverse dynamics of the process. The function is obtained as the solution to integral equation
        \begin{align}
            \begin{split}
                2\mathbf{h}_{\perp} =\;& i \delta E(x,p,\mathbf{h}_{\perp}) \mathbf{g}_{(x,p)}(\mathbf{h}_{\perp}) 
                + \int \frac{\mathrm{d}^2\mathbf{q}_{\perp}}{(2\pi)^2}~\bar{C}(q_\perp)  \nonumber \\
                &\times \Big\{ C_{1} [ \mathbf{g}_{(x,p)}(\mathbf{h}_{\perp}) - \mathbf{g}_{(x,p)}(\mathbf{h}_{\perp} -\mathbf{q}_{\perp}) ] \nonumber\\
                & + \, C_{x} [\mathbf{g}_{(x,p)}(\mathbf{h}_{\perp}) - \mathbf{g}_{(x,p)}(\mathbf{h}_{\perp} -x\mathbf{q}_{\perp}) ]   \nonumber\\
                & + \, C_{1-x} [\mathbf{g}_{(x,p)}(\mathbf{h}_{\perp}) - \mathbf{g}_{(x,p)}(\mathbf{h}_{\perp} -(1{-}x)\mathbf{q}_{\perp}) ]\Big\}.
            \end{split}
            \label{eq:AMY.linear.integral.equation}
        \end{align}
        The factors $C_{1,x,(1-x)}$ are the functions of quadratic Casimir operators, labelled by the momentum fraction of the particle, and given by
        \begin{align}\begin{split}
            C_{1}  =\;&\frac{1}{2} \Big( - C^{R}_{1} + C^{R}_{x} + C^{R}_{1-x} \Big)\;, \\
            C_{x}  =\;&\frac{1}{2} \Big( C^{R}_{1} - C^{R}_{x} + C^{R}_{1-x} \Big)\;,  \\
            C_{1-x}=\;&\frac{1}{2} \Big( C^{R}_{1} + C^{R}_{x} - C^{R}_{1-x} \Big)\;, 
        \end{split}\end{align}
        where $C_{R} = C_{F} = 3/4$ for quarks and $C_{R} = C_{A} = 3$ for gluons. The first term in Eq.~\eqref{eq:AMY.linear.integral.equation} is the energy difference between the initial and final states of the splitting process,
        \begin{equation}
            \delta E(x,p,\mathbf{h}_{\perp}) = \frac{\mathbf{h}_{\perp}^2}{2p\,x(1{-}x)} + \frac{m^2_{\infty,(x)}}{2xp}  + \frac{m^2_{\infty,(1{-}x)}}{2(1{-}x)p} -\frac{m^2_{\infty,(1)}}{2p}.
        \end{equation}
        Here, $m^2_{\infty}$'s are the asymptotic masses of the partons, with their identity indexed by their momentum fraction. With $N_c$ and $N_f$ denoting the number of colors and flavors, respectively, the thermal masses are given by
        \begin{align}
            m^2_{\infty,q} =\;& \frac{4\pi}{3} \alpha_s T^2\nonumber\\
            m^2_{\infty,g} =\;& \frac{2\pi}{3} \left(N_c + \frac{N_f}{2}\right) \alpha_s T^2.
            \label{eq:asymptotic.masses}
        \end{align}
        Finally $\bar{C}(q_\perp)$ is the transverse momentum broadening kernel or the collision kernel with the color factor removed. Its expression at leading order (LO) of the strong coupling is given by~\cite{Arnold:2008vd} 
        \begin{align}
            \bar{C}_{\mathrm{LO}}\left(\mathbf{q}_{\perp}\right) &= \frac{g^2 T^3}{q^2_{\perp}\left(q^2_{\perp} + m^2_D\right)}\int \frac{\mathrm{d}^3p}{\left(2\pi\right)^3} \frac{p-p_z}{p}\times\nonumber
            \\& [2 C_A f_B(p)\bar{f}_B(p')+ 4N_{f} T_f f_F(p)\bar{f}_F(p')]\,,
            \label{eq:transverse.momentum.broadening.kernel.LO}
        \end{align}
        where $f_{F}$ and $f_{B}$ are the Fermi--Dirac and Bose--Einstein distributions, respectively, and the same bar notation is used to signal Pauli blocking or Bose enhancement. The gluon Debye mass, $m_D$, is related to the asymptotic gluon mass of Eq.~\eqref{eq:asymptotic.masses} via
        \begin{align}
            m^2_D =\;& 2\;m^2_{\infty,g}\nonumber\\
                  =\;& \frac{4\pi}{3} \left(N_c + \frac{N_f}{2}\right) \alpha_s T^2.
            \label{eq:gluon.debye.mass}
        \end{align}
        More recently, systematic application of the techniques of \textit{Electrostatic} QCD or \textit{EQCD} have allowed for the next-to-leading order (NLO)~\cite{Caron-Huot:2008zna} and non-perturbative (NP) evaluations~\cite{Moore:2021jwe} of this kernel. In this work we compute the transition rates using the three broadening kernels, with the rate sets referred to by the order of the collision kernel that was used to calculate them. Thus the ``LO'' results in a calculation within a dynamic simulation are computed using rates with the LO collision kernel and so on. 
        
        When computing the new rates and using them in energy loss simulations, the rate equation itself (Eq.~\ref{eq:AMY.linear.integral.equation}), is still leading order and at an implementation level and the radiated partons are taken to be perfectly collinear to the incoming parton. The transverse broadening and separation is then induced by the hard parton's elastic collisions with the thermal medium. This is an approximation, and a more thorough study of high order radiative rates, using the new higher order kernels, would require a careful analysis of radiative events with finite opening angles. Such a study would require a modification to -- or a reorganization of -- the \martini\, framework, as suggested by Ref.~\cite{Dai:2020rlu}. This is a task beyond the scope of this work. Here, we focus on expanding on our work in Ref.~\cite{Yazdi:2022bru} and study the specific effect of the change of the collision kernel on the LO-\amy\, framework when applied to a dynamically expanding medium, with a scale dependent \alphas. 

    \subsection{Collisional Energy Loss}
        Collisional energy loss in \martini\, is provided by $2\rightarrow 2$ LO elastic scattering channels. These are gluon exchange diagrams and preserve the identity of the incoming hard parton. The rates of these processes are computed within kinetic theory. For a process written as $1+2\to 3+4$ where $1$ and $3$ denote the hard incoming and outgoing partons, the rate is given by
        \begin{align}
        	&\frac{d\Gamma_{\mathrm{elas.}}}{d\omega}\left(p,\omega,T\right) =\;\frac{d_2}{(2\pi)^3}\frac{1}{16p^2}\int_0^{p}dq\int^{\infty}_{\frac{q-\omega}{2}}dp_2\,\, \Theta\left(q-|\omega|\right)\nonumber\\
        	&\times\int_0^{2\pi}\frac{d\phi_{p_2q|pq}}{2\pi} |\mathcal{M}|^2 f_2(p_2,T)(1\pm f_4(p_2+\omega,T))\,.
        	\label{eq:differential.scattering.rate}
        \end{align}
        In the above, $d_2$ and $p_2$ denote the degeneracy and the momentum of the incoming medium particle. The momentum of the incoming hard parton is given by $p$ and $\omega$ refers to the energy loss of the hard parton in this process. Furthermore, $f_2$ and $f_4$ give the distribution functions for particles $2$ and $4$, either Bose-Einstein or Fermi-Dirac, depending on the identity of the parton. To screen the divergence in the matrix element from ultra-soft gluon exchange, the Hard-Thermal-Loop (HTL) gluon propagator is used. 

        Elastic scatterings of the jet parton can result in energy subtraction or deposition in the QGP medium. The deposition of energy can result in a jet-induced flow when the exchanged energy from the jet to the medium particle is large enough to ``promote'' a thermal parton to a jet parton. The effect of this event is increased production of relatively softer hadrons around the jet and has been shown to be an important ingredient in reproducing jet substructure observables~\cite{Park:2018acg}. \martini\, accounts for this correlated background by using the \textit{recoil-hole} prescription~\cite{Park:2018acg}. The recoil particle (thermal particle promoted to a jet parton) is taken into the event record and evolved as a jet parton. The hole left in the medium as a result is also added to the event record and allowed to free stream. At the end of the evolution, jet and hole partons are hadronized separately. The hole-hadrons are then subtracted from jet observables at analysis level.
        
    \subsection{Conversion Processes}
        Beyond $2\to 2$ gluon exchange diagrams, a jet parton can experience $2\to 2$ fermion exchange interactions. These processes modify the identity of the incoming jet parton and are dominated by the Mandelstam-$t$ channel. As such, they can be approximated as \textit{collinear conversion} processes where energy loss during the process is assumed minimal and the outgoing hard parton inherits the full energy and momentum of the incoming hard particle. The rates of these process are 
        \begin{align}
            \frac{d\Gamma^{q(\bar{q})\to g}}{d^3k} =\;&\frac{2\pi\,\alpha^2_s}{k}\,T^2 C_{F} \left[\frac{1}{2}\ln{\frac{2kT}{m^2_{\infty,q}}}-0.36149\right]\nonumber\\
            &\times\delta^{(3)}\left(\mathbf{p}-\mathbf{k}\right)\nonumber\\
            =\;&\frac{N^2_c-1}{N_c N_f}\frac{d\Gamma^{g\to q(\bar{q})}}{d^3k}\nonumber\\
            =\;& \frac{\alpha_s}{e^2_f\;\alpha} \frac{d\Gamma^{q(\bar{q})\to\gamma}}{d^3k}\,.
            \label{eq:conversion.rates}
        \end{align}
        

%% file: evolving_sim.tex
\section{Jet energy loss in evolving medium}\label{sec:jet.eloss.evolving.medium}

Here we present an overview of the parameters of the \martini\, framework. This is then followed by a brief discussion of the event-by-event relativistic hydrodynamic simulations used in the calculations of jet energy loss.

    \subsection{Parton Evolution \& MARTINI \ Parameters}\label{sec:martini.framework}
        \martini\, uses \pythia~\cite{Sjostrand:2014zea} in generating the hard collision event. The final, active hard partons in the event record are then evolved in the QGP medium, interacting via the radiative or elastic scattering processes previously discussed in Sec.~\ref{sec:eloss.theory}. The hard partons are evolved individually and are frozen out of the evolution either when they leave the thermal medium or if they can no longer be considered energetic or hard. The former condition is expressed by a temperature cutoff, $T<T_{\mathrm{cut}}$, typically chosen to be above the cross-over temperature $T_{\mathrm{cr}}\approx 150$ MeV. Here, this temperature threshold was fixed to $T_{\mathrm{cut}}=160$ MeV. For the latter freeze-out condition, a momentum cut is used $p<p_{\mathrm{cut}}$. A similar momentum cut is applied to radiated partons in inelastic processes, where the radiated parton is taken into the event record only if its momentum satisfies $k > p_{\mathrm{cut}}$. Similarly, in elastic scattering processes, the hard parton can deposit energy into the medium and promote a thermal parton to a jet parton. These recoil partons are then evolved as regular jet partons. The momentum cut governing their inclusion into the event record is $p > p_{\mathrm{recoil}}$. Here we set these momentum cuts to be equal to each other and proportional to the local temperature, $p_{\mathrm{cut}}=p_{\mathrm{rad}} = p_{\mathrm{recoil}}=4T$. We note that in this study we do not consider jet energy loss in hadronic matter. Given the lower temperatures and consequently lower interaction rates, energy loss in the hadronic gas stage is expected to be subdominant to the QGP phase. Nevertheless, a recent preliminary study of this phenomenon indicated potential for significant contribution to jet quenching~\cite{Elfner:2020men}, particularly for hadrons with low momentum ($p_T\in [2,10]$ GeV). This is (far) below the transverse momenta that we consider here~\footnote{For some jet substructure observables, this range of transverse momenta is relevant. However, in such cases, energy loss via gluon bremsstrahlung in QGP is the dominant relative to energy loss in the hadronic gas, due to significantly higher temperatures in the QGP phase.}.

        In high energy physics phenomenology, the strong coupling is typically treated as a free parameter, to be determined from the data. Once evaluated at a given scale, it is then allowed to run according to renormalization group equations. We follow the same practice here. The most important parameters of \martini\, are those governing its running coupling. They are also the focus of our optimization effort later in this text. The running of the strong coupling in \martini\, is evaluated using the LO pQCD expression, 
            \begin{equation}
                \alpha_s(\mu^2) = \frac{4\pi}{\left(11 -\frac{2}{3} N_f\right)\log{\left(\frac{\mu^2}{\Lambda^2_{\mathrm{QCD}}}\right)}}\,,
                \label{eq:alphas.running.lo.expression}
            \end{equation}
        where $\Lambda_{\mathrm{QCD}} = 200\mathrm{\;MeV}$ is the QCD scale parameter. For each energy loss channel, the renormalization scale is taken to be proportional to the mean transferred transverse momentum
        \begin{equation}
            \mu = \sqrt{\langle p^2_T\rangle}=\begin{cases}
                \kappa_e \sqrt{\hat{q} \lambda_{\mathrm{mfp}}} & \mathrm{elastic\;\&\; conv. \;channels}\\
                \kappa_r \left(\hat{q}p\right)^{1/4} & \mathrm{radiative\;channels}
            \end{cases}.
            \label{eq:alphas.scales}
        \end{equation}
        The proportionality constants, $\kappa_r$ and $\kappa_e$ are taken as free parameters. In the above, $\hat{q}$ is the mean squared exchanged transverse momentum per unit length and given by the second moment of the HTL re-summed elastic scattering rate~\cite{Arnold:2008vd} 
        \begin{equation}
            \frac{d\Gamma_{\mathrm{elas.}}}{d^2\mathbf{q}_{\perp}} = \frac{C_{R}}{(2\pi)^2}\frac{4\pi \alpha_{s,0} m^2_D T}{\mathbf{q}^2_{\perp}(\mathbf{q}^2_{\perp} + m^2_D)}\,,
            \label{eq:elastic.scattering.rate.running.coupling}
        \end{equation}
        where $m^2_D$ is the squared Debye mass, Eq.~\eqref{eq:gluon.debye.mass}, and $C_{R}$ the quadratic Casimir operator of the jet parton. The expression for $\hat{q}$, then, is given by
        \begin{align}
            \hat{q} =\;& \int^{q_{\mathrm{max}}} d^{2}\mathbf{q}_{\perp} \mathbf{q}^2_{\perp} \frac{d\Gamma_{\mathrm{elas.}}}{d^2\mathbf{q}_{\perp}}\nonumber\\
            =\;&  C_{R}\alpha_{s,0}m^2_D\log{\left(1+\frac{q^2_{\mathrm{max}}}{m^2_D}\right)}.
        \end{align}
        $\alpha_{s,0}$ is the third parameter of the running coupling in \martini\ and it is the scale at which the Debye mass is evaluated. The upper bound on the $\mathbf{q}_{\perp}$ integration is set to $q_{\mathrm{max}}=\sqrt{6pT}$ with $p$ denoting the momentum of the incoming hard parton. The mean free path is the inverse of the total elastic scattering rate, given by
        \begin{equation}
            (\lambda_{\mathrm{mfp}})^{-1} = \int_{q_{\mathrm{min}}}^{q_{\mathrm{max}}} d^{2}\mathbf{q}_{\perp}\frac{d\Gamma_{\mathrm{elas.}}}{d^2\mathbf{q}_{\perp}}\,,
        \end{equation}
        where $q_{\mathrm{min}}=0.05T$ is chosen to be consistent with the elastic scattering rate tables. 

        During the calculations, both channels cut off $\alpha_s$ at $\alpha_{s,\mathrm{max}}=0.42$ while only the elastic channel places a lower bound on the strong coupling, $\alpha_{s,\mathrm{min}}=0.15$. 

       Thus the three main parameters of \martini\, are $\kappa_r, \kappa_e$ and $\alpha_{s,0}$, which control the running of the strong coupling. Among these three, the value of the running coupling is least sensitive to $\alpha_{s,0}$ and as such we fix its value to $\alpha_{s,0}=0.3$. The other two parameters, $\kappa_r, \kappa_e$, are then treated as the free parameters that are to be tuned by data. 

        In this work, we modify the above picture of hard parton generation and evolution in \martini\, framework to also account for a ``shower formation'' time, post hard scattering event. In order to study this effect, and inspired by Ref.~\cite{Zhang:2022ctd}, we assign a formation time to each outgoing hard parton by summing over the splitting time of its parent partons,
            \begin{equation}
                \tau_{\mathrm{form.}} = \sum_i \tau_{\mathrm{form.}, i}.
                \label{eq:form.time.sum}
            \end{equation}
        In the above, $\tau_{\mathrm{form.}, i}$ is given by
            \begin{equation}
                \tau_{\mathrm{form.}, i} = \frac{2E_i\;x_i(1-x_i)}{k^2_{\perp, i}}, 
            \end{equation}
        where $E_i$ is the energy of the parent, $x_i$ and $1-x_i$ the energy fractions taken by the outgoing particles and $k_{\perp, i}$ is the transverse momentum of the outgoing partons relative to the parent. Thus each parton is only allowed to interact with the local thermal medium if the current time $\tau$ is greater than both the formation time of the particle and the start time of the hydro, $\tau > \max{\left(\tau_0, \tau_{\mathrm{form.}}\right)}$. In the terminology of Ref.~\cite{Zhang:2022ctd}, we study Setup 1 (without formation time) and Setup 3 (with formation time from multiple splittings).

    \subsection{Soft Medium}\label{sec:hydro.sims}
        The event-by-event hydrodynamic simulations of the heavy ion events used in this work are provided by \music~\cite{Schenke:2010nt}, a $(3+1)D$ relativistic viscous hydrodynamic simulator which solves the conservation equation of energy-momentum tensor and the relaxation equations of shear stress tensor and bulk viscous pressure~\cite{Molnar:2013lta, Denicol:2014vaa}. Here, the stress-tensor for fluid dynamics simulations are initialized at $\tau=0.4$ fm$/c$ by two-dimensional \ipg\, calculations~\cite{Schenke:2012hg,Schenke:2012wb}. After this initialization, the system is evolved using the HotQCD Collaboration equation of state~\cite{Bazavov_2014}, a constant shear viscosity to entropy density $\eta/s = 0.13$, and the same parametrization for the bulk viscosity as in Ref.~\cite{Ryu:2015vwa}. The values for hydro start time as well as transport coefficients used in this work are similar to those found in the recent state-of-the-art Bayesian study of the soft sector in Ref.~\cite{Heffernan:2023gye}. 

%% file: section_formation_time.tex
\subsection{Importance of shower formation-time}\label{sec:fit.and.formtime}

    In Sec.~\ref{sec:martini.framework}, we indicated that the \martini\, framework was modified to account for the formation-time of final state parton shower. The effect of the inclusion of formation-time in the high virtuality part of the parton shower is studied by comparing \martini\, simulations with and without formation-time, both against each other as well as against experimental observations. In simulations without formation-time in the high-virtuality stage, the fully developed \pythia\, parton shower is fed directly to \martini\, for propagation in the evolving QGP medium while in those with formation-time, each outgoing hard parton is assigned a formation-time according to Eq.~\eqref{eq:form.time.sum}.  
            
    \begin{table}
        \centering
        \begin{tabular}{l l l}
        \toprule 
            Parameter & Value & Note\\ 
        \midrule
            $N_c$ & $3$ & number of colors\\
            $\Lambda_\mathrm{QCD}$ & $0.2~\text{GeV}$ & Eq.~\protect{\eqref{eq:alphas.running.lo.expression}}\\
            $p_{\mathrm{cut}}$ & $4T$ & universal cut: $p_{\mathrm{cut}}, p_{\mathrm{recoil}}, p_{\mathrm{eloss}}$\\
            $N_f$ & $3$ & number of flavors\\
            $\alpha_{s,0}$ & $0.3$ & Eq.~\protect{\eqref{eq:elastic.scattering.rate.running.coupling}}\\
            $T_{\mathrm{frz}}$ & $0.16$ GeV & jet freeze-out temperature\\
        \bottomrule
        \end{tabular}
        \caption{Summary of the fixed \martini\, parameters used in this study.}
        \label{tab:fixed.params}
    \end{table}
            
    The important parameters of \martini, as mentioned in Sec.~\ref{sec:martini.framework}, are those that govern the running of the strong coupling. Thus we use these parameters and their determination or optimization as our instruments in studying the effect of the formation-time in the high-virtuality shower. The targets of this parameter tuning are $\kappa_r$ and $\kappa_e$, the coefficients of the renormalization scale of the strong coupling in radiative and elastic scattering channels, respectively. A group of fifty pairs of $\kappa=(\kappa_r,\kappa_e)$ are sampled from the interval $\kappa_{r,e}\in [0, 15]$. In this section, we focus on the LO rate set, since NLO and NP rates yield similar results in a dynamic simulation. For convenience, we summarize all other fixed parameters in Tab.~\ref{tab:fixed.params}. Simulations resulting from all $\kappa$ pairs are presented in the figures below, before the final presentation of best fit results. This serves the purpose of illuminating how changing the $\kappa$ values changes the behaviour of the model in calculating the observables considered here. Furthermore, it makes explicit the importance of including high-virtuality shower formation-time. 
            
    \begin{figure*}
        \begin{subfigure}[t]{0.45\linewidth}
            \centering
            \includegraphics[width=\linewidth]{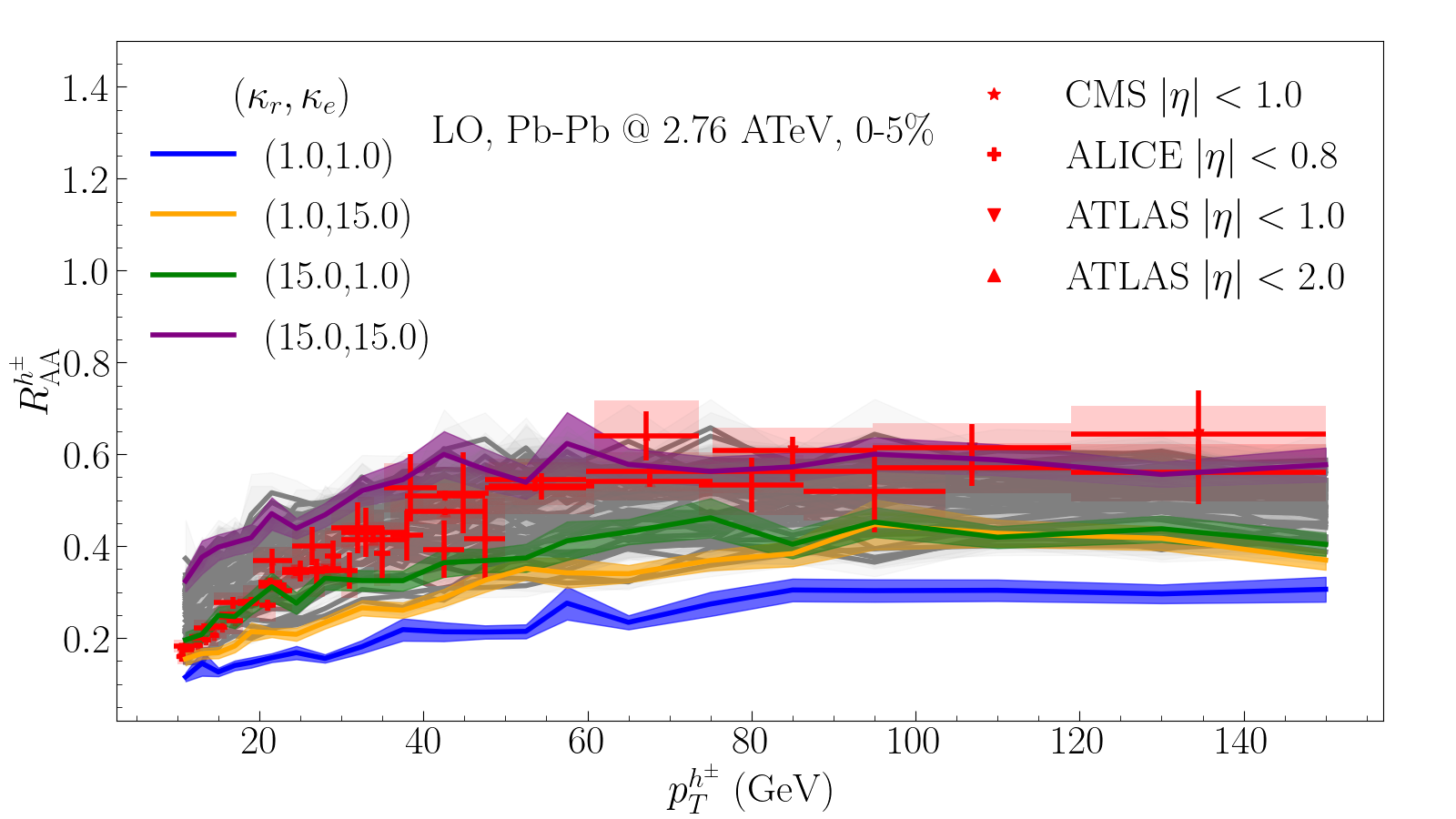}
            \caption{Charged hadron nuclear modification factor at midrapidity ($|\eta^{h^{\pm}}|<1$). Data points are experimental observations charged hadron nuclear modification factors at midrapidity from CMS~\cite{CMS:2012aa}, ATLAS~\cite{ATLAS:2015qmb} and ALICE~\cite{ALICE:2012aqc} Collaborations.}
            \label{fig:fit.coverage.charged.RAA.LO.no.ftime}
        \end{subfigure}
        \begin{subfigure}[t]{0.45\linewidth}
            \centering
            \includegraphics[width=\linewidth]{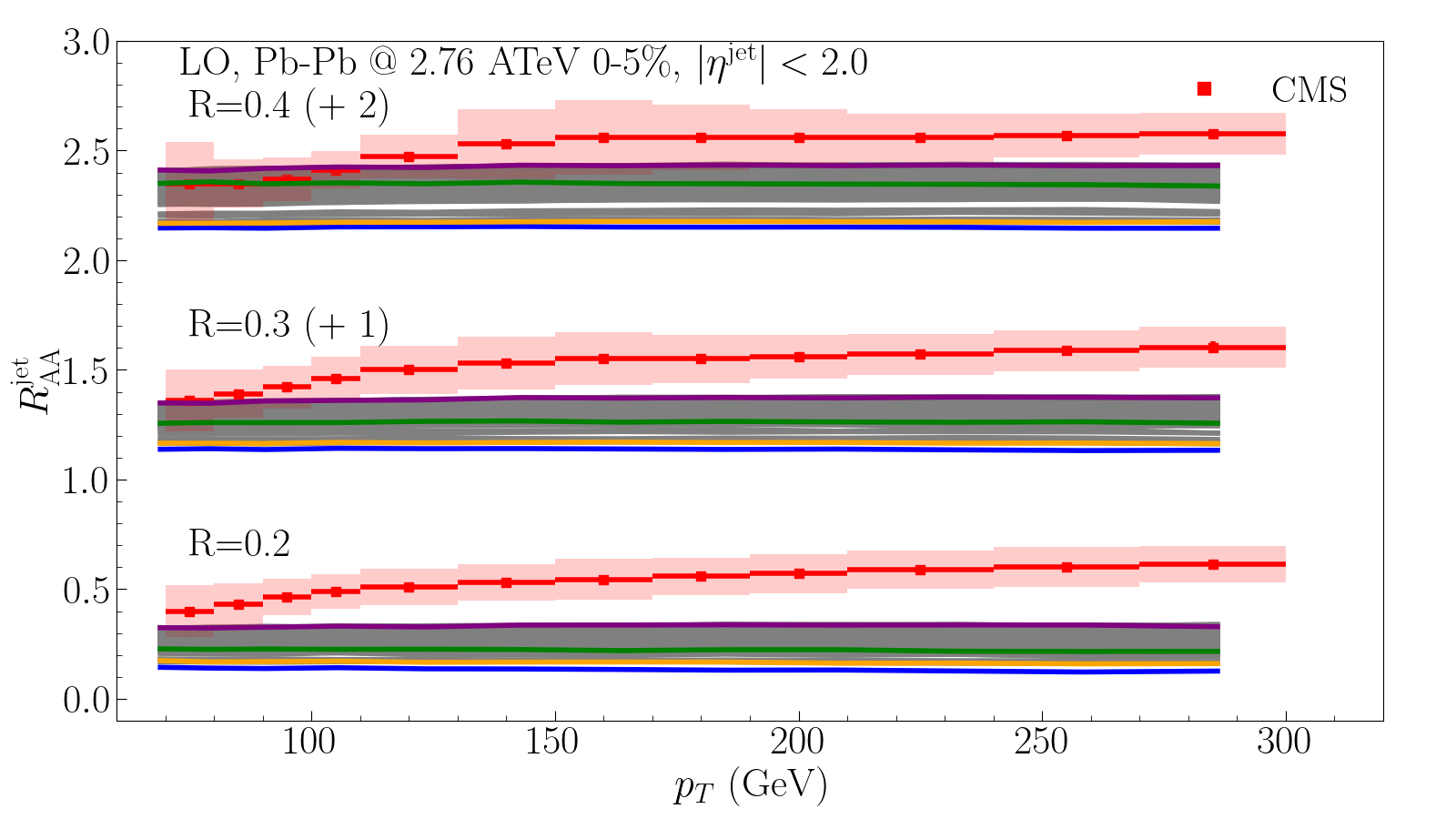}
            \caption{Nuclear modification factor of jets. The jets are clustered for pseudorapidity window $|\eta|<2$ and three different cone radii, $R\in[0.2,0.3,0.4]$. For better separation and clarity, a constant value is added to $R=0.3$ and $0.4$. Data from the CMS Collaboration~\cite{CMS:2016uxf}.}
            \label{fig:fit.coverage.jet.raa.no.ftime}
        \end{subfigure}
        \caption{Nuclear modification factor of (a)~charged hadron (left) and (b)~jet (right) for \pbpb\, collisions at $\sqrt{s}=2.76$ ATeV and $0$-$5\%$ centrality. These do not include a formation-time in the high-virtuality parton shower. The bounding sets where $\kappa_{r,e}=1$ or $15$ are given in color while other runs with different choices of these parameters are shown in gray. Experimental observations are shown in red. The jets are clustered from the same events as the charged hadron results, using the anti-$k_T$ algorithm.}
        \label{fig:no.form.time.charged.hads.and.jets}
    \end{figure*}
    
    We first consider the charged hadron nuclear modification factor, 
    \begin{equation}
        R^{h^{\pm}}_{\mathrm{AA}} = \frac{dN^{h^{\pm}}_{\mathrm{AA}}/dp_Td\eta}{N_{\mathrm{bin.}}dN^{h^{\pm}}_{\mathrm{pp}}/dp_Td\eta}\;,
         \label{eq:charged.hadron.raa.cov.no.ftime}
    \end{equation}
   where $N_{\mathrm{bin.}}$ is the number of binary collisions. The results of the charged hadron \raa\, calculations for the different $\kappa$ sets are shown in Fig.~\ref{fig:fit.coverage.charged.RAA.LO.no.ftime}. 
   Another feature of this plot, other than the observable similarity of the $R_{\mathrm{AA}}$ from different $\kappa$ pairs, is the significant quenching even for $\kappa=(15, 15)$. Given that the $\kappa$ values multiply the renormalization scale, a larger $\kappa_{r,e}$ implies a smaller $\alpha_s$ for the respective channel. Therefore, while we have good overall agreement with the data and an observation of a slight curvature as $\kappa$ values increase, the observed quenching is more than what one would expect. 

   In order to potentially distinguish between different $\kappa$ sets, one can consider jets and jet substructure. These are three-dimensional objects, and expected to be more sensitive to the interplay of the radiative and elastic channels of energy loss. Similar to the charged hadron \raa\, in Eq.~\eqref{eq:charged.hadron.raa.cov.no.ftime}, jet nuclear modification factor is given by

    \begin{equation}
        R^{\mathrm{jet}}_{\mathrm{AA}} = \frac{dN^{\mathrm{jet}}_{\mathrm{AA}}/dp_Td\eta}{N_{\mathrm{bin.}}dN^{\mathrm{jet}}_{\mathrm{pp}}/dp_Td\eta}\;.
        \label{eq:jet.raa}
    \end{equation}
    Throughout this work, jets are clustered using the anti-$k_T$ algorithm~\cite{Cacciari:2008gp} for different jet cone radii $R$ ,
    \begin{equation}
        R = \sqrt{\Delta \eta^2 + \Delta \phi^2},
    \end{equation}
    where $\Delta \eta\equiv \eta_i - \eta_j$ and $\Delta \phi\equiv \phi_i-\phi_j$ are the difference in pseudorapidity and azimuthal angle (in momentum space) between different hard partons ($i$ and $j$) during the clustering process~\cite{Cacciari:2008gp}.

    Using the same events as Fig.~\ref{fig:fit.coverage.charged.RAA.LO.no.ftime}, we can construct the inclusive jet nuclear modification factor in Fig,~\ref{fig:fit.coverage.jet.raa.no.ftime}. 
    While the charged hadron \raa\, results were able to readily reproduce the data across a vast set of parameters, the jet \raa\, results indicate significant over-quenching: none of the $\kappa$ pairs are able to match the data. In this model configuration, the over-quenching is indicative of long evolution time. By considering a low-virtuality model and having an instantaneous parton shower, all partons coming out of the hard interaction point are confronted with an evolving medium and see the whole evolution history. 
    
    Given the poor performance of jet \raa\, relative to the data and its unsuitability as a fit target for simulations without formation-time, we consider the jet-shape ratio, given by
    \begin{align}
        R^{\rho}_{\mathrm{AA}} =& \frac{\rho_{\mathrm{AA}}(r)}{\rho_{\mathrm{pp}}(r)}\,,\label{eq:jet.shape.ratio}\\
        \rho(r) \equiv& \frac{N_\mathrm{norm}}{N_\mathrm{jet}} \frac{\sum_\mathrm{jets} \sum_{r \in [r_\mathrm{min},r_\mathrm{max})} {p_{T}^{\mathrm{trk}}}/{p_{T}^{\mathrm{jet}}}}{r_\mathrm{max} - r_\mathrm{min}}\,,\nonumber\\
        r =& \sqrt{(\phi_\mathrm{trk} - \phi_\mathrm{jet})^2 + (y_\mathrm{trk}-y_\mathrm{jet})^2}\,. \nonumber
    \end{align}
    
\begin{figure}
        \centering
        \includegraphics[width=\linewidth]{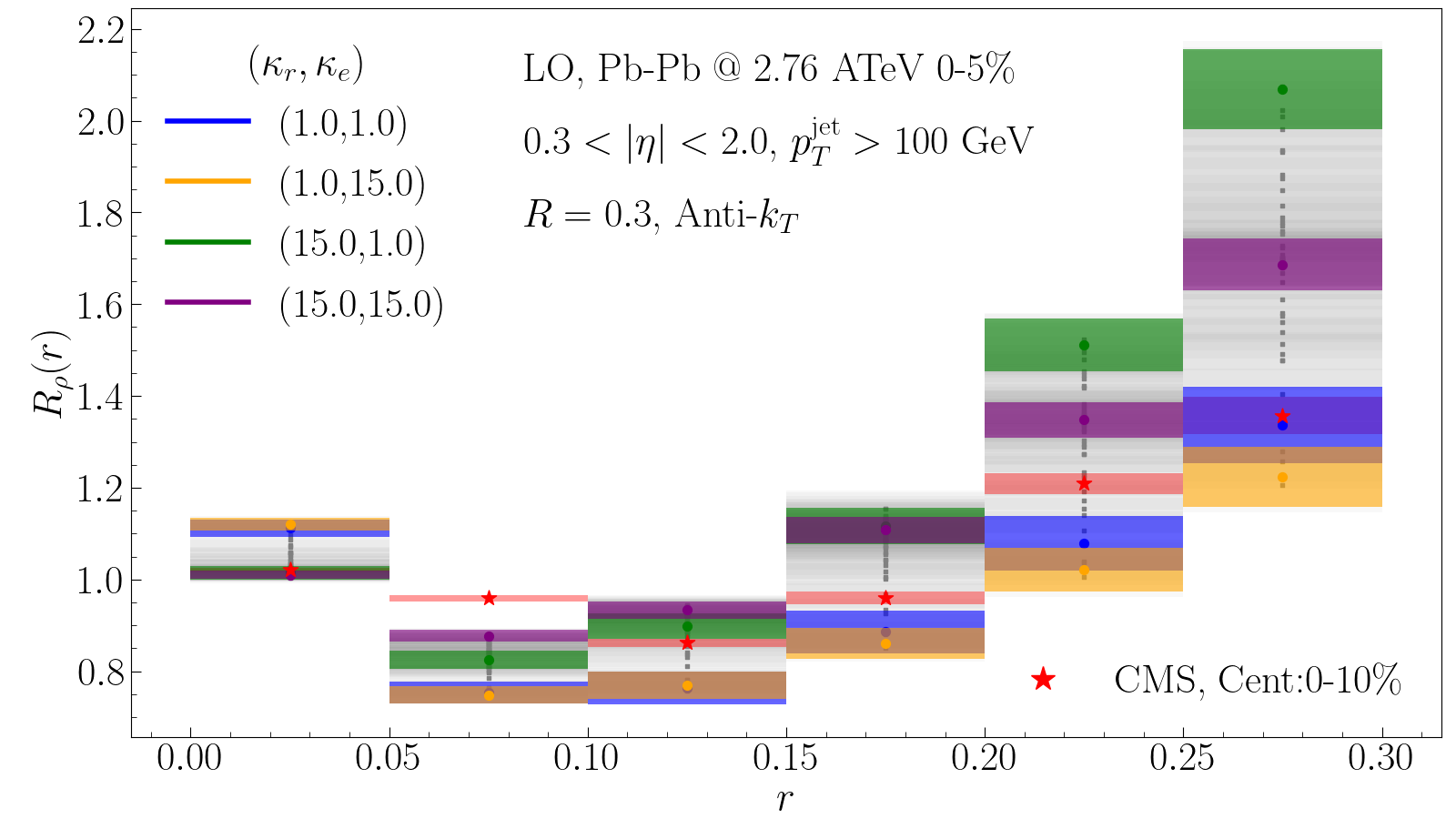}
        \caption{Jet shape ratio for \pbpb\, collisions at $2.76$ ATeV, for $0$-$5\%$ centrality, using the same events as Fig.~\ref{fig:no.form.time.charged.hads.and.jets}. The box around theory calculations denotes the statistical uncertainty. A transverse momentum cut is placed on the jets, $p^{\mathrm{jet}}_T>100$ GeV as well as the charged tracks $p^{\mathrm{trk}}_T>1$ GeV. Jets are clustered with $R=0.3$ cone radius for $0.3 < |\eta| < 2.0$. The data, shown as red stars, belong to the $0$-$10\%$ centrality class of \pbpb\, collisions, and are taken from the CMS Collaboration~\cite{CMS:2013lhm}. The boxes denote statistical uncertainty.}
        \label{fig:fit.coverage.shape.LO.no.ftime}
    \end{figure}

    Fig.~\ref{fig:fit.coverage.shape.LO.no.ftime} shows the results of the jet shape analysis from the same events. We see that that the calculations from different $\kappa$ sets provide good coverage of the data. It should be noted that as a result of the poor performance of the model against the jet \raa\, data, the jet shape observable is being constructed for the \textit{wrong} jet population. Thus, in an over-quenched or over-evolved simulation, one can easily recover charged hadron \raa\, and the jets will \textit{look} correct, but their actual population will be wrong and too much energy would have migrated down to the lower energy modes.   
    
    Here, for our purposes, we can still continue with the optimization procedure. One can now consider the ``instantaneous parton shower'' approximation as a ``long evolution time'' study of \amy\, rates in a dynamic medium, rather than a QGP brick. The parameter optimization for this model will use charge hadron \raa\, as well as jet shape ratio results of Fig.~\ref{fig:fit.coverage.shape.LO.no.ftime} as targets of the fit. In this way, we are able to use observables that are sensitive to the different energy loss channels and can potentially distinguish between different $\kappa$ pairs in Fig.~\ref{fig:fit.coverage.charged.RAA.LO.no.ftime}. As stated before, jet \raa\, itself is not used for the optimization of the models without formation-time.

    \begin{figure*}
        \begin{subfigure}[t]{0.45\linewidth}
            \centering
            \includegraphics[width=\linewidth]{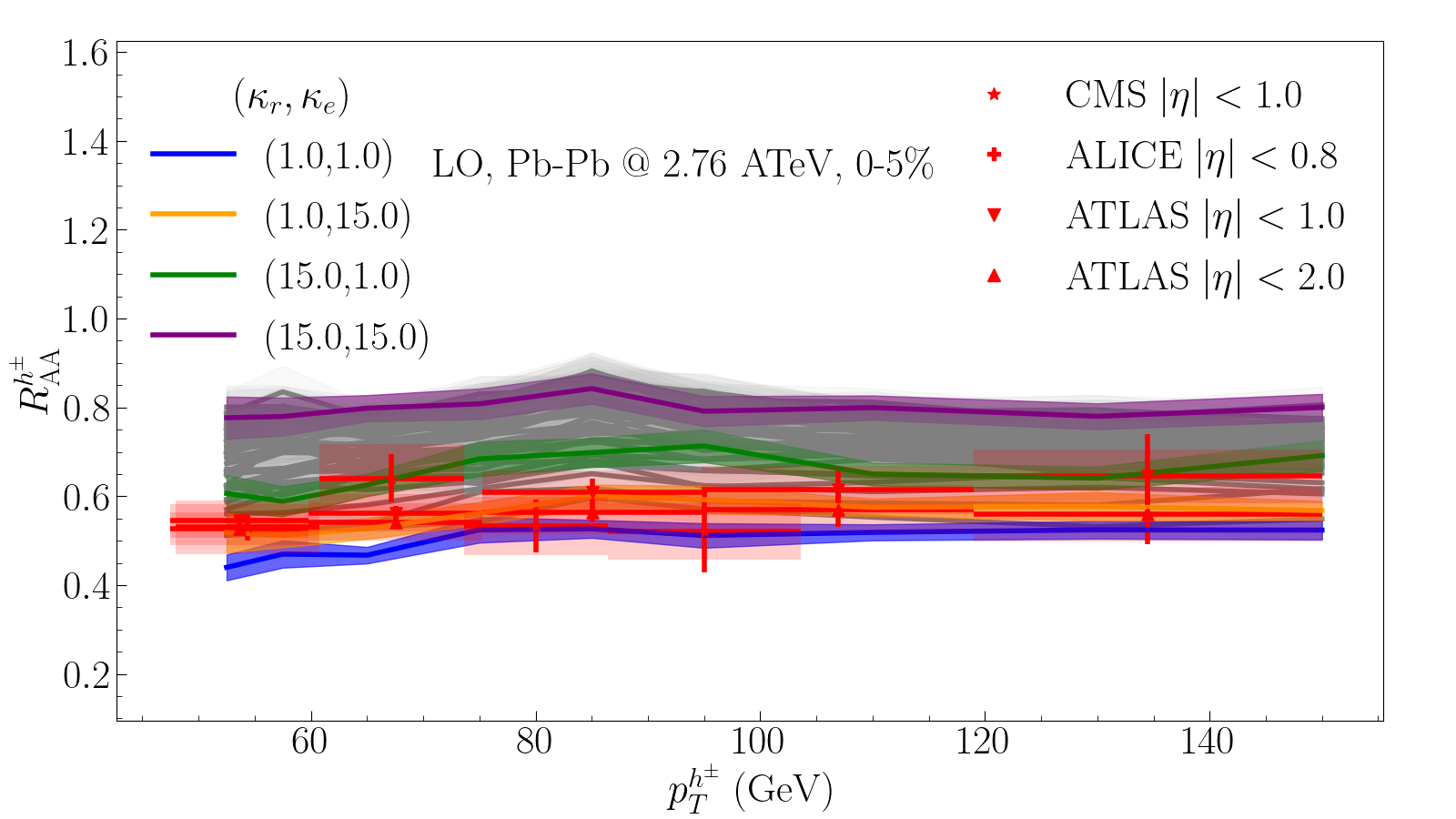}
            \caption{Same as \autoref{fig:fit.coverage.charged.RAA.LO.no.ftime} but with a time-delayed parton shower.}
            \label{fig:fit.coverage.charged.RAA.LO.with.ftime}
        \end{subfigure}
        \begin{subfigure}[t]{0.45\linewidth}
            \centering
            \includegraphics[width=\linewidth]{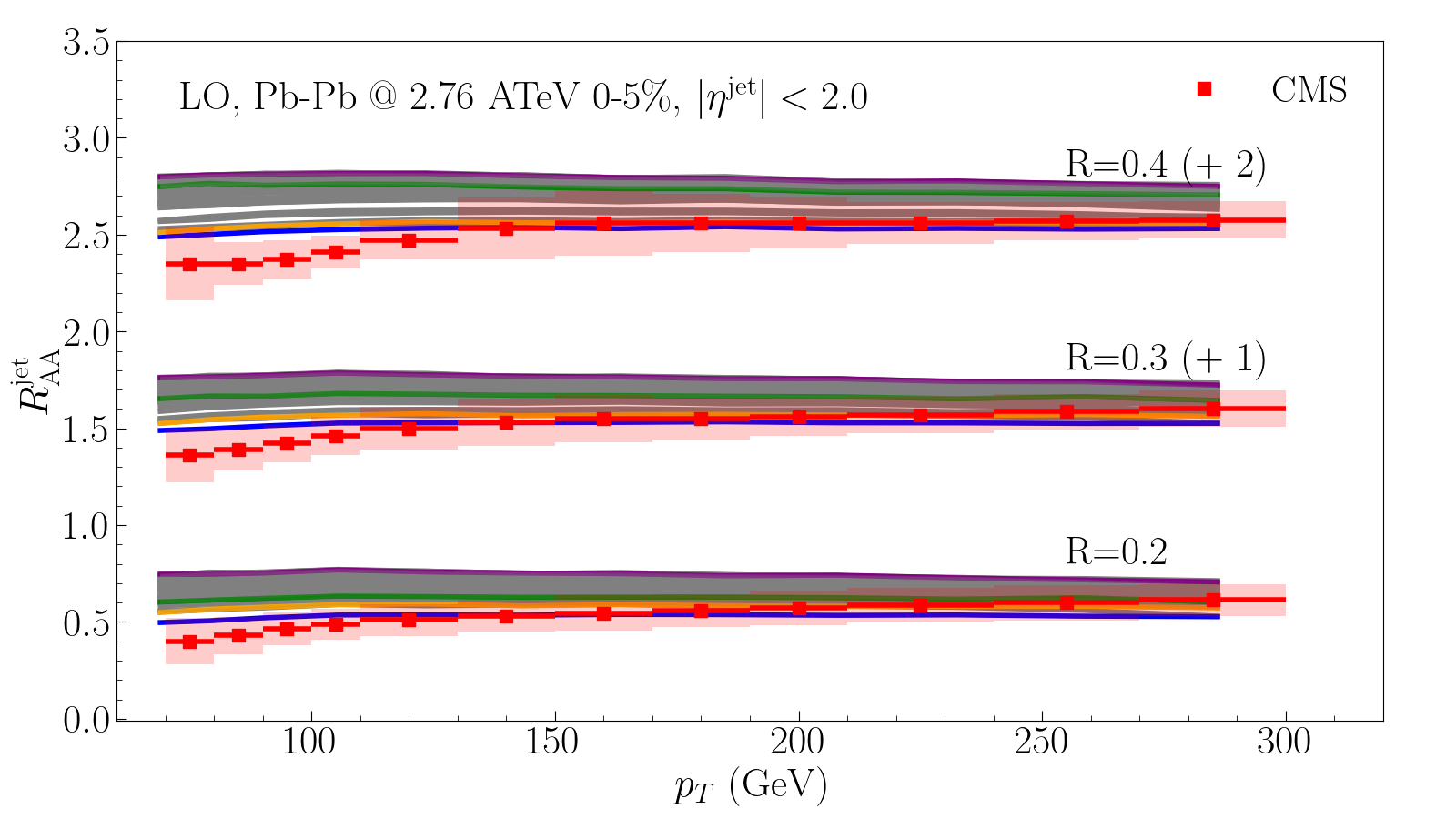}
            \caption{Same as \autoref{fig:fit.coverage.jet.raa.no.ftime} but with a time-delayed parton shower.}
            \label{fig:fit.coverage.jet.raa.with.ftime}
        \end{subfigure}
        \caption{Nuclear modification factor of (a)~charged hadron (left) and (b)~jet (right) for \pbpb\, collisions at $\sqrt{s}=2.76$ ATeV and $0$-$5\%$ centrality. Unlike Fig.~\ref{fig:no.form.time.charged.hads.and.jets}, here the high-virtuality parton shower includes formation-time. The jets used in the Fig.~(b) are clustered from the same events as Fig.~(a).}
        \label{fig:with.form.time.charged.hads.and.jets}
    \end{figure*}
    
    For the second model under study, we introduce a formation-time in the high virtuality shower, the effect of the inclusion of which is immediately clear in Fig.~\ref{fig:fit.coverage.charged.RAA.LO.with.ftime}. Whereas previously the simulation could barely come above the charged hadron \raa\, data, even for the largest pairs of $\kappa$ (corresponding to smaller values of \alphas), now we can span the data more easily. 
            
    This observation continues in Fig.~\ref{fig:fit.coverage.jet.raa.with.ftime}, where the effect of the delayed parton shower is much more striking. The over-quenching that was previously seen in Fig.~\ref{fig:fit.coverage.jet.raa.no.ftime} is entirely gone and the simulations can easily match the magnitude of the observed \raa\, for a range of $\kappa$ pairs. 
    \begin{figure}
        \centering
        \includegraphics[width=\linewidth]{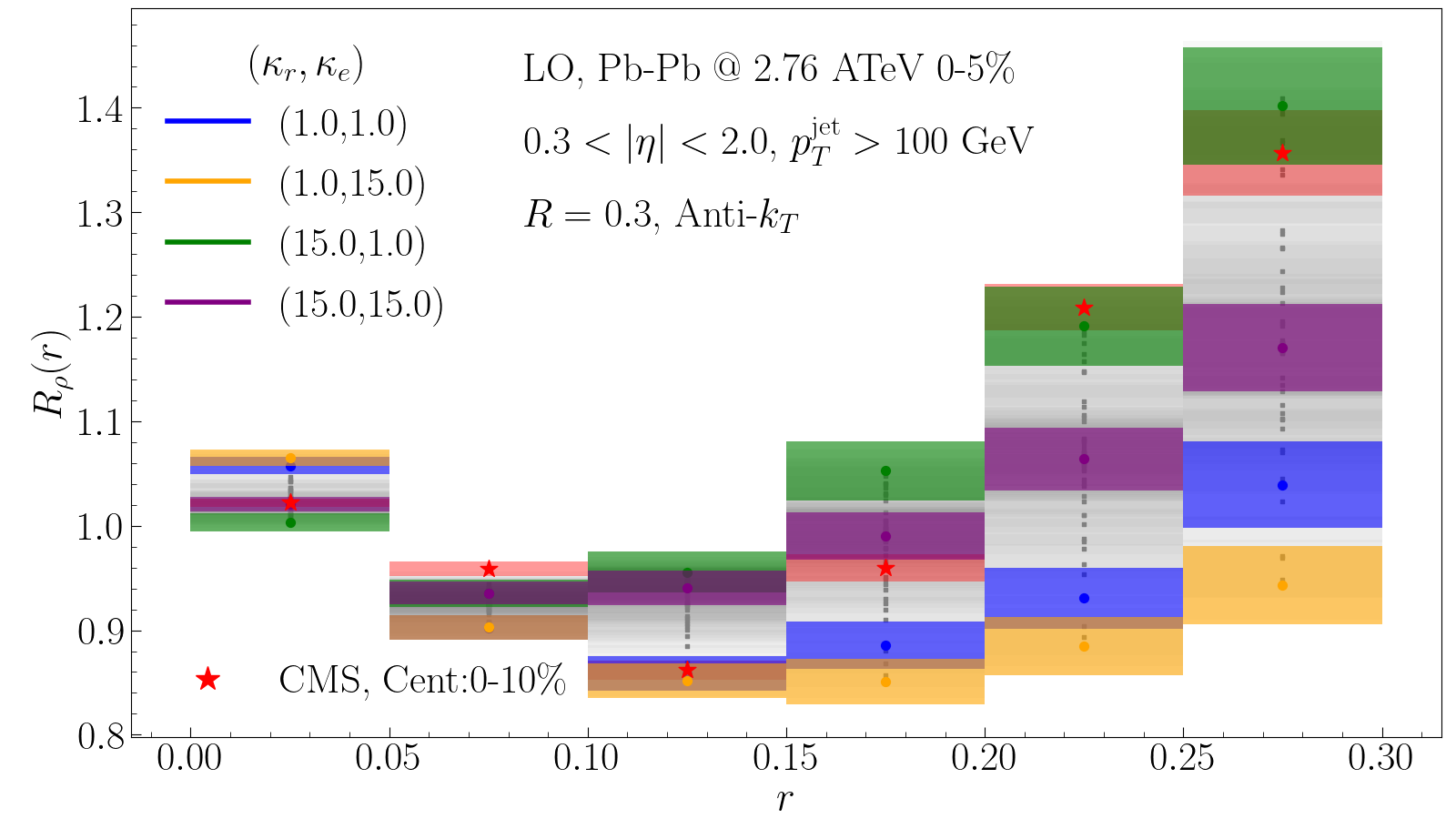}
        \caption{Jet shape ratio for simulations with formation-time in the high-virtuality shower. The jets are clustered from the same events as Fig.~\ref{fig:with.form.time.charged.hads.and.jets}. Same kinematic cuts as Fig.~\ref{fig:fit.coverage.shape.LO.no.ftime} are applied.}
        \label{fig:fit.coverage.shape.LO.with.ftime}
    \end{figure}
            
    Inclusion of a formation-time for the partons of the shower, however, changes the shape of the jets, as seen in Fig.~\ref{fig:fit.coverage.shape.LO.with.ftime}, and brings them closer to jets in \pp\, collisions. Comparing Fig.~\ref{fig:fit.coverage.shape.LO.no.ftime} and Fig.~\ref{fig:fit.coverage.shape.LO.with.ftime}, jet shape from each $\kappa$ set in the latter is noticeably more flat than the equivalent set in the former. By including a formation-time, some shower partons from the initial hard scattering only begin interacting with the thermal medium at later times where lower temperatures in the QGP medium mean lower energy loss rates. In some cases, a shower parton may have a formation-time that is longer than the lifetime of the plasma and therefore never interacts with the medium at all. 
             
    \subsubsection{Parameter Optimization}\label{sec:fit.proc.and.res}
            
        \begin{table*}
            \centering
            \begin{tabular}{l l}
            \toprule 
                Observable & Note\\ 
            \midrule
                Charged Hadron \raa\, & Used for both models, $|\eta|<1.0$\\
                Jet Shape Ratio ($R_{\rho}$) & Used for both models, $0.3<|\eta|<2.0$, $p^{\mathrm{jet}}_T > 100$ GeV\\
                Jet \raa\, & Delayed parton shower model only, $|\eta|<2.0$, $R\in[0.2,0.3,0.4]$\\
            \bottomrule
            \end{tabular}
            \caption{Observables used in the tuning of the models.}
            \label{tab:tuning.obs}
        \end{table*}
            
        From here on, the two models above will be referred to as ``no formation-time'' and ``with formation-time''. The fit procedure is identical for both. The target observables for each model are summarized in Tab.~\ref{tab:tuning.obs}. 

        For each model and observable, we calculate the $\chi^2$,
            \begin{equation}\label{eq:chi.squared.def}
                \chi^2/\mathrm{d.o.f} = \frac{1}{\sum_i 1}\sum_{i} \frac{\left(y_{\mathrm{expt},i}-y_{\mathrm{theor},i}\right)^2}{\sum_s \sigma^2_{s,i}},
            \end{equation}
        where the index $i$ runs over the data points of the observable and index $s$ covers the uncertainties, systematic and statistical, that are associated with each point. In order to optimize the parameters of the model, $\kappa_r$ and $\kappa_e$, we use Gaussian Process Regression~\cite{rasmussen2005gaussian}, and obtain the optimized parameters by the condition
            \begin{equation}
                (\kappa^{\mathrm{opt}}_r,\kappa^{\mathrm{opt}}_{e}) = \mathrm{argmin}\; G_{\chi^2}.
            \end{equation}
        In the above, $G_{\chi^2}$ is the Gaussian regressor trained on the calculated $\chi^2$ values of Eq.~\eqref{eq:chi.squared.def}.
            \begin{table}
                \centering
                \begin{tabular}{c  c  c}
                \toprule
                    Parameter & No $\tau_{\mathrm{form.}}$& With $\tau_{\mathrm{form.}}$ \\
                    \midrule
                     $\kappa_r$ & 2.0 & 1.0\\
                     $\kappa_e$ & 8.6 & 2.5\\
                    \bottomrule
                \end{tabular}
                \caption{Optimal parameters for the LO-rate set, with and without a shower formation-time. The parameters were tuned using the observables in Tab.~\ref{tab:tuning.obs}. See the full list of $\kappa$ values for the new collision kernels as well as the LO kernel in Tab.~\ref{tab:all.kappa.params}.}
                \label{tab:lo.kappa.params}
            \end{table}
        \autoref{tab:lo.kappa.params} shows the results of the parameters. As one would expect, without incorporating formation-time into the parton shower, the optimal $(\kappa_r, \kappa_e)$ pair is found to be relatively large in order to reduce the value of the resulting \alphas. Delaying the parton shower by including a formation-time results in the reduction of the values of the parameters and therefore an increase in the calculated \alphas. 
            \begin{figure}
                \centering
                \includegraphics[width=\linewidth]{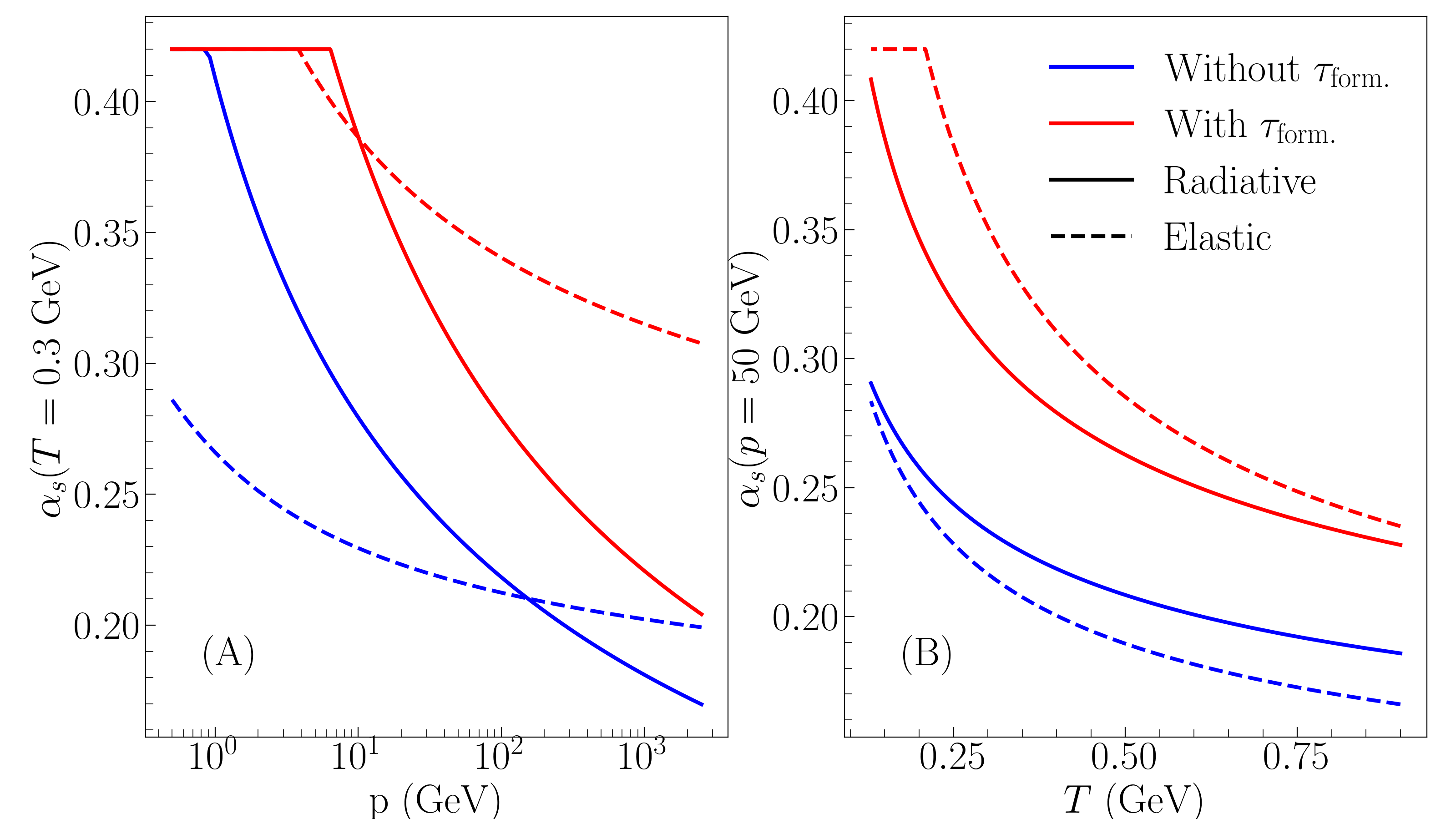}
                \caption{Comparison of the value of the running coupling using the optimal parameters show in Tab.~\ref{tab:lo.kappa.params} for the two energy loss channels. Solid and dashed lines denote radiative and elastic channels respectively. Figure A (left) shows \alphas\, at a representative temperature ($T=0.3\;$ GeV) as a function of parton momentum. Figure B (right) fixes the jet momentum at $p=50$ GeV. See text for details.}
                \label{fig:compare.alphas.LO}
            \end{figure}
        This is summarized in Fig.~\ref{fig:compare.alphas.LO} where the running coupling is plotted as a function of momentum and temperature for the two energy loss channels. It is evident from the figure that the inclusion of a formation-time for the parton shower can significantly alter the running \alphas. 
        
        \subsubsection{Realistic evolution with optimized parameters} \label{sec:runs.with.fitted.params}
        
            \begin{figure}
                \centering
                \includegraphics[width=\linewidth]{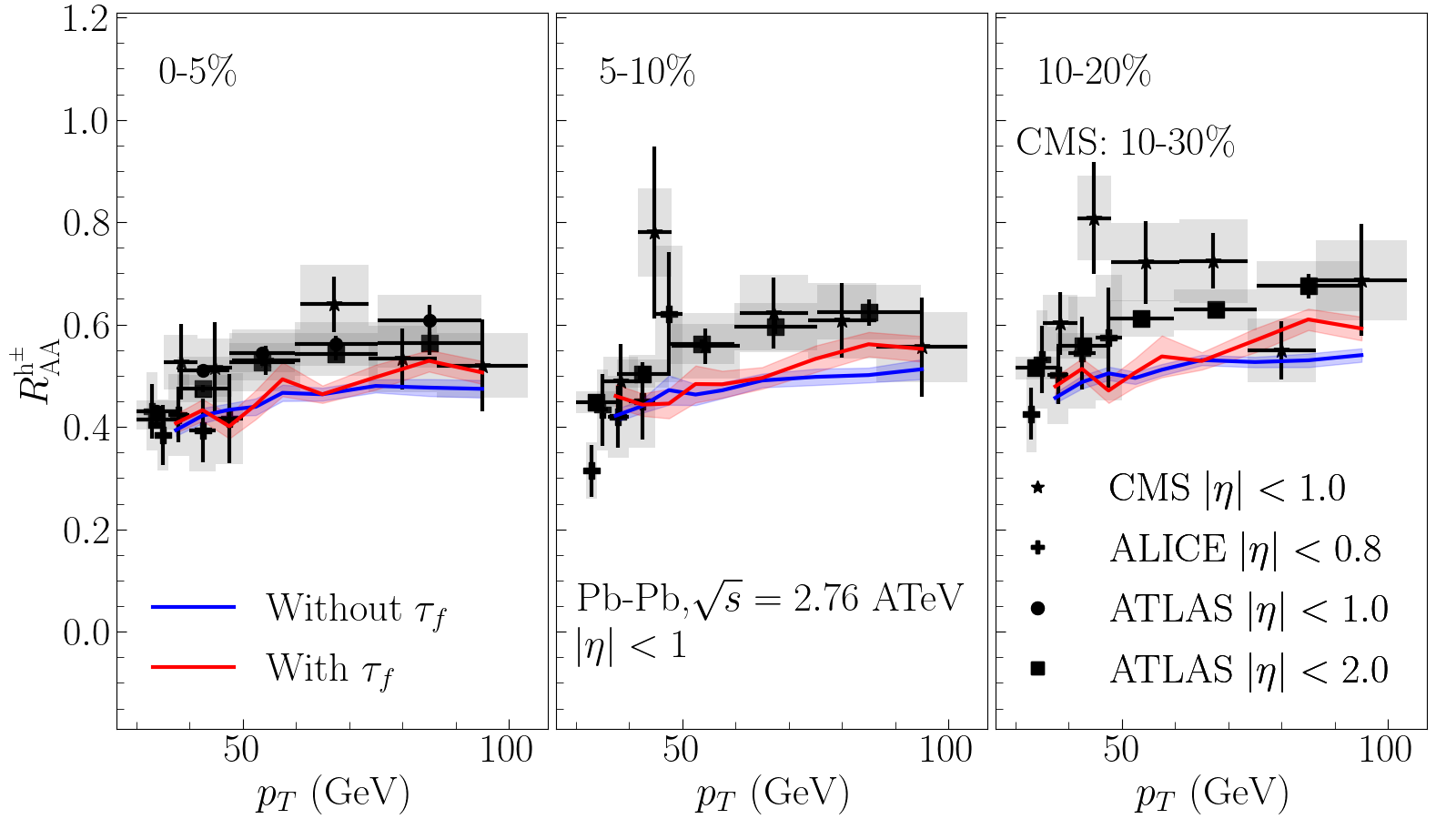}
                \caption{Inclusive charged hadron \raa\, for \pbpb\, collisions at $\sqrt{s}=2.76 $ ATeV and three centrality classes, $0$-$5$, $5$-$10$ and $10$-$20\%$, using the new optimized values for $\kappa_r$ and $\kappa_e$. We see very similar comparison to data whether or not formation-time is included in the shower. Data from the ALICE~\cite{ALICE:2012aqc}, ATLAS~\cite{ATLAS:2015qmb} and CMS~\cite{CMS:2012aa} Collaborations.}
                \label{fig:compare.charged.raa.formation.time.final}
            \end{figure}
            Using the new optimized parameters for the running coupling, we can compare the models directly. We return to the hadronic and jet observables that were introduced before and refer the interested reader to App.~\ref{sec:pythia.guns} for a look at the effect of formation on the evolution of the hard parton spectrum. Fig.~\ref{fig:compare.charged.raa.formation.time.final} shows the results of a dynamical simulations, with the new parameters, for charged hadron \raa\, in three centrality classes. The $0$-$5\%$ centrality class was used in the tuning of the parameters while the other two centralities are predictions. While the model without formation-time is visually below the one with formation-time, the two curves are mostly overlapping each other (within statistical uncertainties). The agreement between the two models in this observable as well is their performance relative to the data across three centrality classes is quite good. This is a consequence of the inclusive nature of the charged hadron \raa\, which makes it a robust observable and quite simple to replicate.   

            \begin{figure}
                \centering
                \includegraphics[width=\linewidth]{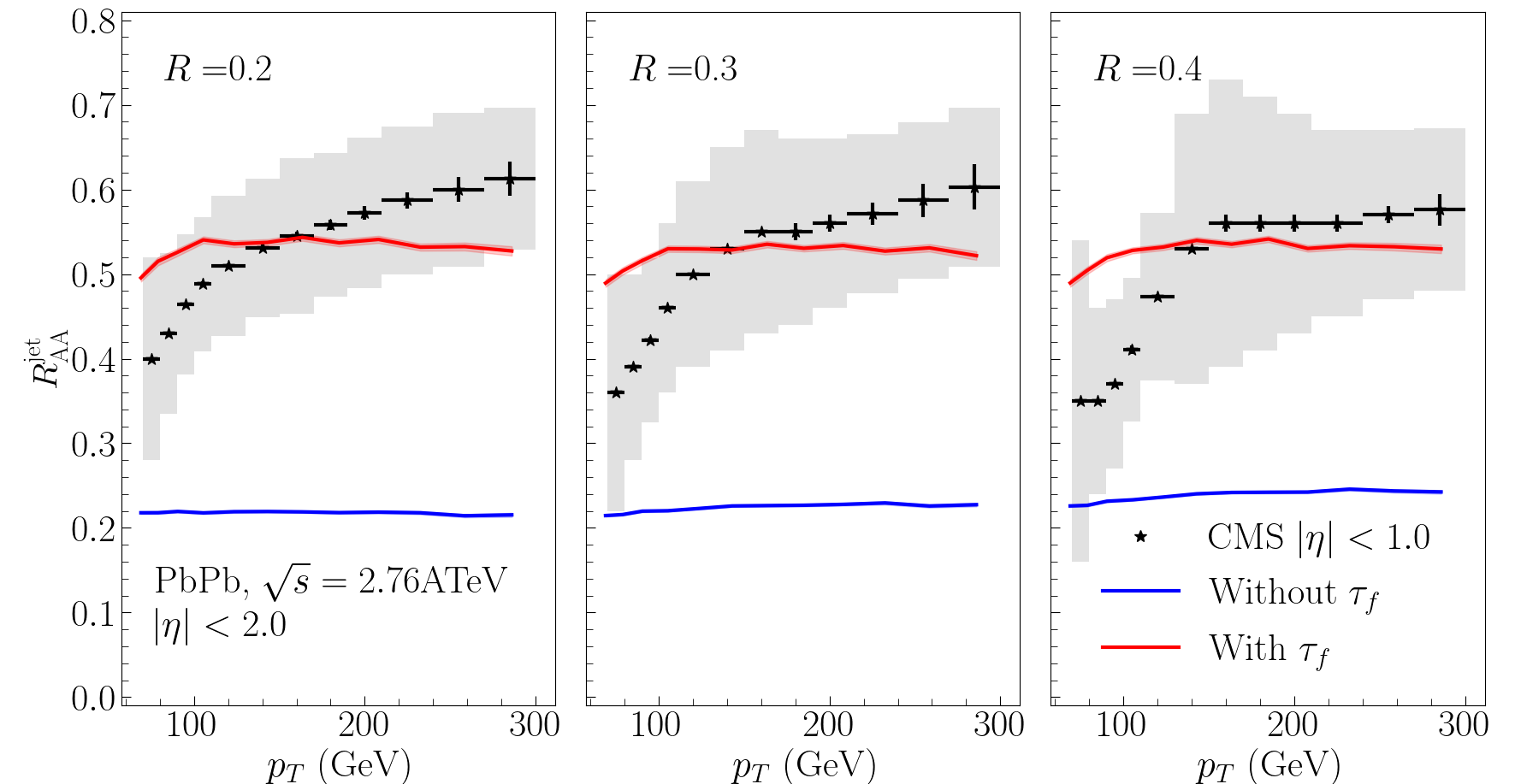}
                \caption{Inclusive jet \raa\, for \pbpb\, collisions at $2.76$ ATeV and $0$-$5\%$ centrality. The jets are clustered using the anti-$k_T$ algorithm from the same events as Fig.~\ref{fig:compare.charged.raa.formation.time.final}. The relative jump in the \raa\, calculation as a result of formation-time in the parton shower is clear. Data from the CMS Collaboration~\cite{CMS:2016uxf}.}
                \label{fig:compare.jet.raa.formation.time.final}
            \end{figure}

            The jet nuclear modification factor shown in Fig.~\ref{fig:compare.jet.raa.formation.time.final} provides the most visually striking difference between a delayed parton shower and an instantaneous one. It should be noted that the presented data was used in the tuning of the parameters of the models. The figure shows inclusive jets clustered using the same events that generated Fig.~\ref{fig:compare.charged.raa.formation.time.final} and while charged hadrons were seemingly insensitive to properties of the parton shower, jet \raa\, is more than doubled and in much better agreement with the data. 

            \begin{figure}
                \centering
                \includegraphics[width=\linewidth]{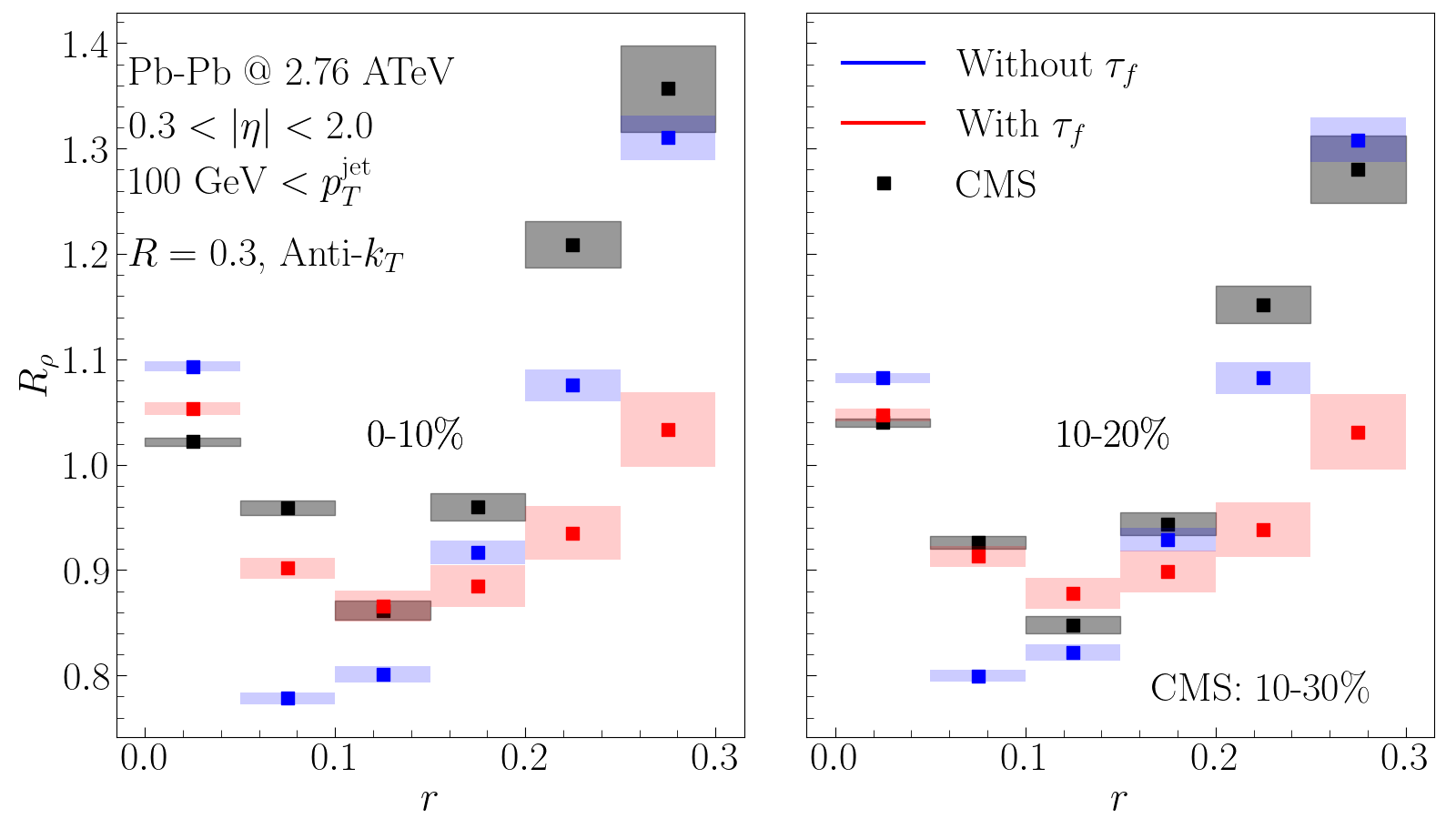}
                \caption{Jet shape ratio using the same events as Fig.~\ref{fig:compare.charged.raa.formation.time.final} and Fig.~\ref{fig:compare.jet.raa.formation.time.final}. Jets are clustered using anti-$k_T$ algorithm with $p^{\mathrm{jet}}_T>100$ GeV, a momentum cut on the charged tracks $p^{\mathrm{trk}}_{T}>1$ GeV and $0.3<|\eta|<2.0$. The $0$-$10\%$ centrality class was used in the tuning of the parameters and the $10$-$20\%$ calculation is a prediction though the data is for $10$-$30\%$ centrality. Data taken from the CMS Collaboration~\cite{CMS:2013lhm}.}
                \label{fig:compare.jet.shape.formation.time.final}
            \end{figure}

            \begin{figure}
                \centering
                \includegraphics[width=\linewidth]{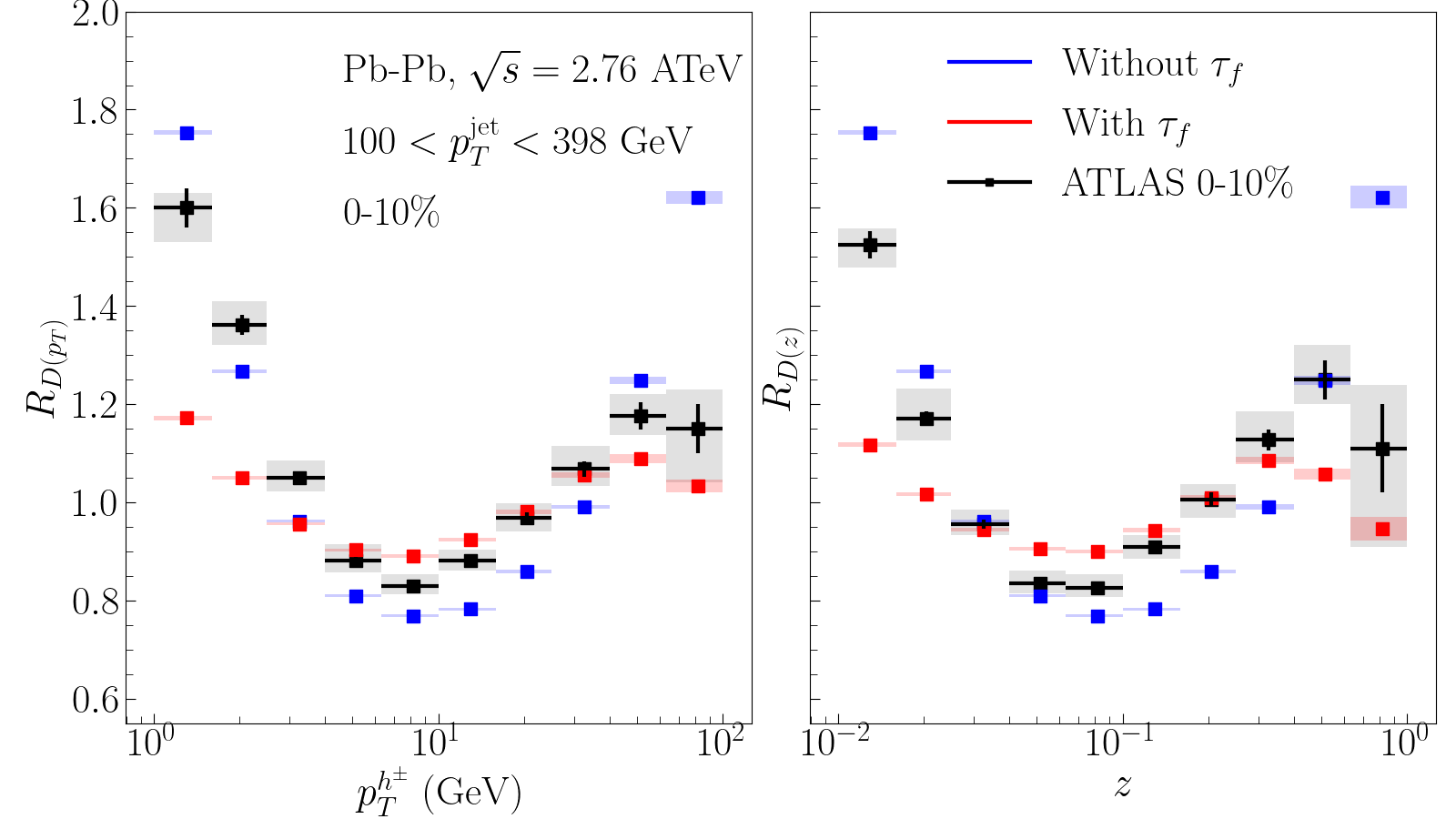}
                \caption{Comparison of the jet fragmentation function ratios for the $0$-$10\%$ centrality class of \pbpb\, collisions at $2.76$ ATeV. The results are from the same events as Figs.~\ref{fig:compare.charged.raa.formation.time.final}, ~\ref{fig:compare.jet.raa.formation.time.final} and ~\ref{fig:compare.jet.shape.formation.time.final}. The jets are clustered using the anti-$k_T$ algorithm, and the fragmentation function ratios are calculated for jets with $100<p_T<398$ GeV and at midrapidity $|\eta|<2.1$. Data from the ATLAS Collaboration~\cite{ATLAS:2017nre}.}
                \label{fig:compare.jet.FF.formation.time.final}
            \end{figure}

            Jet shape ratio results are presented in Fig.~\ref{fig:compare.jet.shape.formation.time.final}, where the $0$-$10\%$ centrality data was used in the tuning and the $10$-$20\%$ data is the prediction using the new parameters. The model without a formation-time in the shower exhibits much better agreement in the outer annuli of the jet ($0.15 \lesssim r$) while the model with formation-time does a better job at capturing the population of charged hadrons closer to the jet axis ($r\lesssim 0.15$). The latter model generally exhibits a jet shape that is closer to the proton-proton baseline, with only $\sim 14\%$ deviation for intermediate values of $r$. This is indicative of the inflexibility of our implementation of a formation-time for the parton shower. While a time delayed parton shower is more physical than an instantaneous parton shower, or a parton shower that has placed all out-going partons on the mass shell before $\tau_{\mathrm{hydro}}$, it is still an approximation. The evolving, highly virtual partons can still interact with the dynamic thermal medium and receive elastic scatterings. This would induce more emission and therefore modify the time it takes for the hard partons to approach the mass shell. Thus, some partons may in fact begin to interact earlier with the medium than what is allowed in our simulations with formation-time for the parton shower. The extra interaction time would then allow for more radiative and elastic scatterings which would improve the performance of the model with respect to the jet shape ratio data, particularly for $r>0.16$. It is emphasized that the jet shape ratio calculated for the model with formation time is perhaps closer to the ``correct'' or ``real'' population of these jets than the model without formation time. As previously discussed, the jet population in the no formation time model is over-quenching and represents a ``long evolution'' time calculation. It is interesting that while the leading charged hadrons and the smaller annuli in the $\eta-\phi$ plane are missed by this model, the soft tail (or outer annuli) of jet shape look as they should when compared to data. 

            We can flesh out this point further by considering an observable that was not used in the tuning process of either model, which provides complementary information to jet shape: jet fragmentation function ratios. Jet fragmentation function (FF) is given by
            \begin{align}\begin{split}\label{eq:FF.z}
    	    D(z)_{z \in [z_\mathrm{min},z_\mathrm{max})} 
              \equiv\;& 
            \frac{\sum_\mathrm{jets} \sum_{z \in [z_\mathrm{min},z_\mathrm{max})}1}{N_\mathrm{jet} \; (z_\mathrm{max} - z_\mathrm{min})}
    	 \, ,
            \end{split}\end{align}
            where $N_\mathrm{jet}$ is the total number of jets within the selected kinematic region, and $z$ is the charged hadron momentum fraction along the direction of the jet,
            \begin{align}
                z \equiv \frac{\mathbf{p}_\mathrm{jet} \cdot \mathbf{p}_\mathrm{trk}}{\mathbf{p}_\mathrm{jet} \cdot \mathbf{p}_\mathrm{jet}} \,.
            \end{align}
            This observable can also be defined with respect to the hadron $p_T$
            \begin{align}\begin{split}\label{eq:FF.pT}
    	    D(p_T)_{p_T \in [p_{T}^{\mathrm{min}}, p_{T}^{\mathrm{max}})} 
                \equiv\;& 
	        \frac{\sum_\mathrm{jets} \sum_{p_{T,\mathrm{trk}} \in [p_{T}^{\mathrm{min}},p_{T}^{\mathrm{max}})}1}{N_\mathrm{jet} \; (p_{T,\mathrm{max}} - p_{T,\mathrm{min}})}\,.
            \end{split}\end{align}
            Thus Eqs.~\eqref{eq:FF.z} and~\eqref{eq:FF.pT} use the same information though they weight it differently. Jet FF ratios are, then, given by
            \begin{align}
                R_{D(p_T)} =\;& \frac{D(p_T)_{\mathrm{AA}}}{D(p_T)_{\mathrm{pp}}}\nonumber\\
                R_{D(z)} =\;& \frac{D(z)_{\mathrm{AA}}}{D(z)_{\mathrm{pp}}}.
            \end{align}
            \autoref{fig:compare.jet.FF.formation.time.final} shows the result of the calculation of jet FF ratios for the $0$-$10\%$ centrality class and completes the story.~The model without a shower formation-time matches (at least qualitatively) the data for $z<0.5$ or $p_T<50$ GeV while missing the leading charged hadron. The model exhibits a larger than expected number of charged hadrons in the highest $p_T$ or $z$ bin in Fig.~\ref{fig:compare.jet.FF.formation.time.final} but this is due to over-quenching of even higher energy jets, and it systematically under estimates the intermediate $p_T$ hadrons, as their energy has been shifted even lower. 

            By considering the results above, we conclude that it is not possible to achieve a simultaneous description of charged hadron and jet data in realistic simulations with a parton shower that is fully developed before the start of the hydrodynamic phase. This argument is also supported by results of simulation using \lbt\, in calculations of parton energy loss in Ref.~\cite{Zhang:2022ctd}. Jet substructure observables also benefit from the addition of a formation-time in the high virtuality shower. However, we find that a complete description of jet substructure, soft and hard modes simultaneously, would require a more dynamic approach to the development of the high virtuality shower. As an example, and to study the effect of a more realistically developing parton shower, we refer the interested reader to App.~\ref{sec:compare.jetscape.matter} for a few illustrative comparisons of \martini\, with and without shower formation time and multi-stage simulations using \matter\  \cite{Majumder:2013re,Majumder:2014gda,Cao:2017qpx}, a model of jet energy loss in the high virtuality stage which handles the final state shower and accounts for its spacetime dependence.

%% file: section_high_order_kernels.tex
\subsection{Effect of higher order collision kernels} \label{sec:higher.order.kernels}
    In the previous section, we established the importance of the inclusion of formation time in the development of the high-virtuality shower. In this section, we study the effect of the higher order collision kernels on jet propagation in a QGP medium. As mentioned in Sec.~\ref{sec:rad.energy.loss.theory}, these kernels are used in the \amy\, framework to create new inelastic splittings rates in a QGP medium. These are then referred to by the collision kernel that was used to generate them: ``LO rate'' for rates using the LO kernel and so on. 
   
    In Ref.~\cite{Yazdi:2022bru} we studied the effect of using the new splitting rates on jet energy loss in QGP bricks of different lengths as well as in a dynamically expanding medium. When considering jet energy loss in the dynamically expanding medium, the value of the strong coupling in the elastic and radiative channels was fixed to a constant (i.e. no scale dependence). This allowed for a straightforward analysis of simulations with different rate sets, as the only difference between them would be the rate set itself. However, this is an idealized condition and one that we relax in this study. 

    \begin{table}
        \centering
        \begin{tabular}{c  c  c}
            Rate set & \multicolumn{2}{c}{$\left(\kappa_r,\kappa_e\right)$}\\
            \midrule 
            & No formation time & With formation time\\
            \midrule
             LO  &  $\left(2.0,8.6\right)$  & $\left(1.0,2.5\right)$\\
             NLO &  $\left(4.4,11.6\right)$ & $\left(1.6,6.8\right)$\\
             NP  &  $\left(2.8,9.8\right)$  & $\left(1.4,3.8\right)$\\
            \bottomrule
        \end{tabular}
        \caption{Optimal parameter sets for the running coupling (Eq.~\ref{eq:alphas.scales}) of the three rate sets. Parameters are given for simulations both with and without a delayed parton shower. Information of Table~\ref{tab:lo.kappa.params} is presented here again for convenience and clarity.}
        \label{tab:all.kappa.params}
    \end{table}
        
    Based on the discussion in Sec.~\ref{sec:fit.and.formtime}, we study the new rates by introducing scale dependence to the strong coupling for jet energy loss. The parameters are independently tuned for each rate set, according to the procedure outlined in Sec.~\ref{sec:fit.proc.and.res}. Furthermore, we present simulations \textit{with formation time} in the final state parton shower. This is to allow for a more ``realistic'' calculation, using the present setup. We refer the interested reader to App.~\ref{sec:long.short.formtime.kernels} for a brief comparison between simulations with and without shower formation time in a dynamically expanding medium. App.~\ref{sec:pythia.guns} shows the effect of using different rates on the evolving hard parton distributions, without the effects of hadronization.
    
    \begin{figure}
        \centering
        \includegraphics[width=\linewidth]{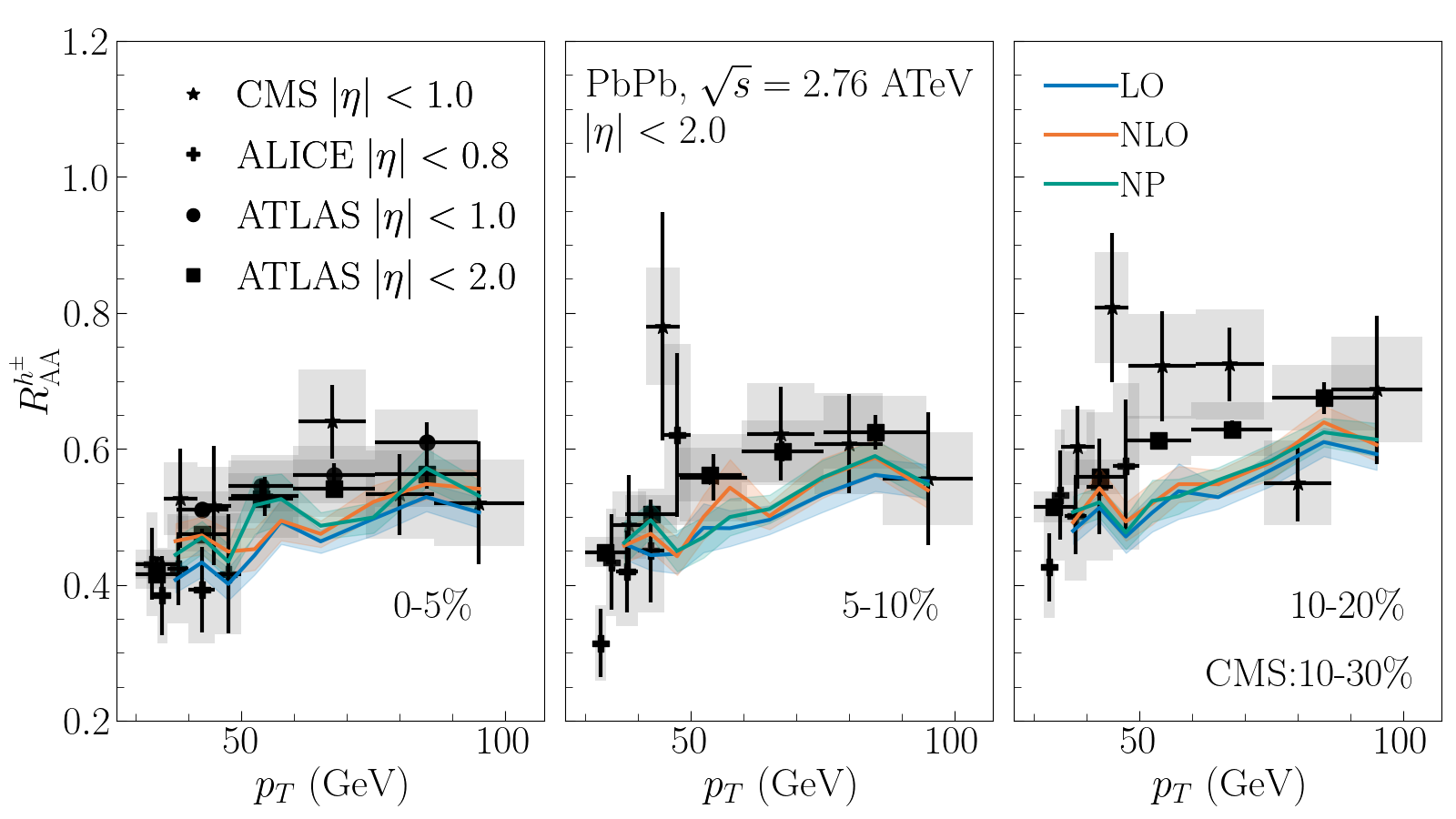}
        \caption{Charged hadron nuclear modification calculated using different collision kernels in the radiative rates at midrapidity ($|\eta|<1.0$) and three centrality classes of \pbpb\, collisions at $2.76$ ATeV center-of-mass collision energy. Simulations include formation time in the final state, high-virtuality parton shower. Data from the ALICE~\cite{ALICE:2012aqc}, ATLAS~\cite{ATLAS:2015qmb} and CMS~\cite{CMS:2012aa} Collaborations.}
        \label{fig:coll.kernels.charged.hadron.raa}
    \end{figure}
    
    \autoref{fig:coll.kernels.charged.hadron.raa} shows the calculation of charged hadron \raa, using the optimized parameters of Table~\ref{tab:all.kappa.params}. The $0$-$5\%$ centrality data was used in the tuning of the parameters of the three models. The other centrality classes are predictions. We observe good agreement between the theory results and the data across all three centralities. Furthermore, much like the fixed \alphas\, results of Ref.~\cite{Yazdi:2022bru}, the charged hadron \raa\, curves of different models seem to have collapsed on top of each other, indicating the insensitivity of this observable to the details of the energy loss calculation. 

    \begin{figure*}
        \centering
        \includegraphics[width=\linewidth]{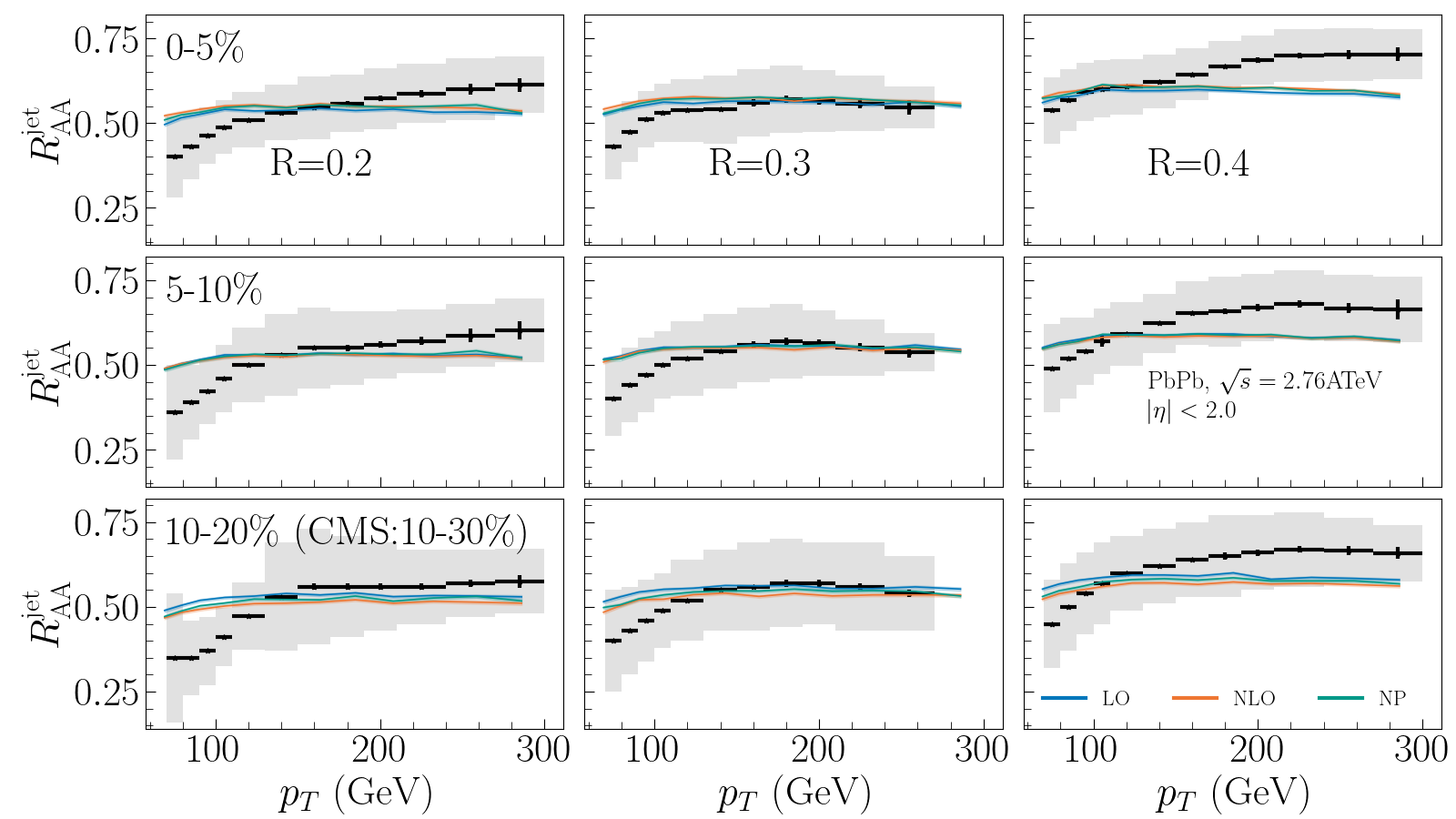}
        \caption{Inclusive jet \raa\, results, using the same events as \autoref{fig:coll.kernels.charged.hadron.raa}. Jets are clustered using the anti-$k_T$ algorithm, for three centrality classes (rows) and three jet cone radii (columns). Data from the CMS Collaboration~\cite{CMS:2016uxf}.}
        \label{fig:coll.kernels.jet.raa}
    \end{figure*}

    Jet \raa\, results using the same events mostly maintain this picture. \autoref{fig:coll.kernels.jet.raa} shows the calculation of inclusive jet \raa\, for the three centrality classes and jet cone radii. The theory calculations compare favorably with the data and appear to also match each other. However, we can observe a weak centrality dependence, where the curves begin to separate from each other when going from $0$-$5$\% to $10$-$20\%$ centrality class. This indicates a slightly different path-length dependence in the three combined elastic and inelastic rate sets. The simulations are seeded by the same hard scattering event and parton showers, and see the same hydrodynamic history. The path length difference, then, stems from the interplay of the elastic scattering channel which uses the same rates but has a different scale dependence (due to different $\kappa_{e}$ for each simulation) and the radiative channels which have different rates (for different collision kernels) and different $\kappa_r$. 
    
    Finally, we can turn our attention to more differential observables, which are more sensitive to the interplay of the elastic and radiative channels. Jet substructure observables are what we turn to for this task.
    \begin{figure}
        \centering
        \includegraphics[width=\linewidth]{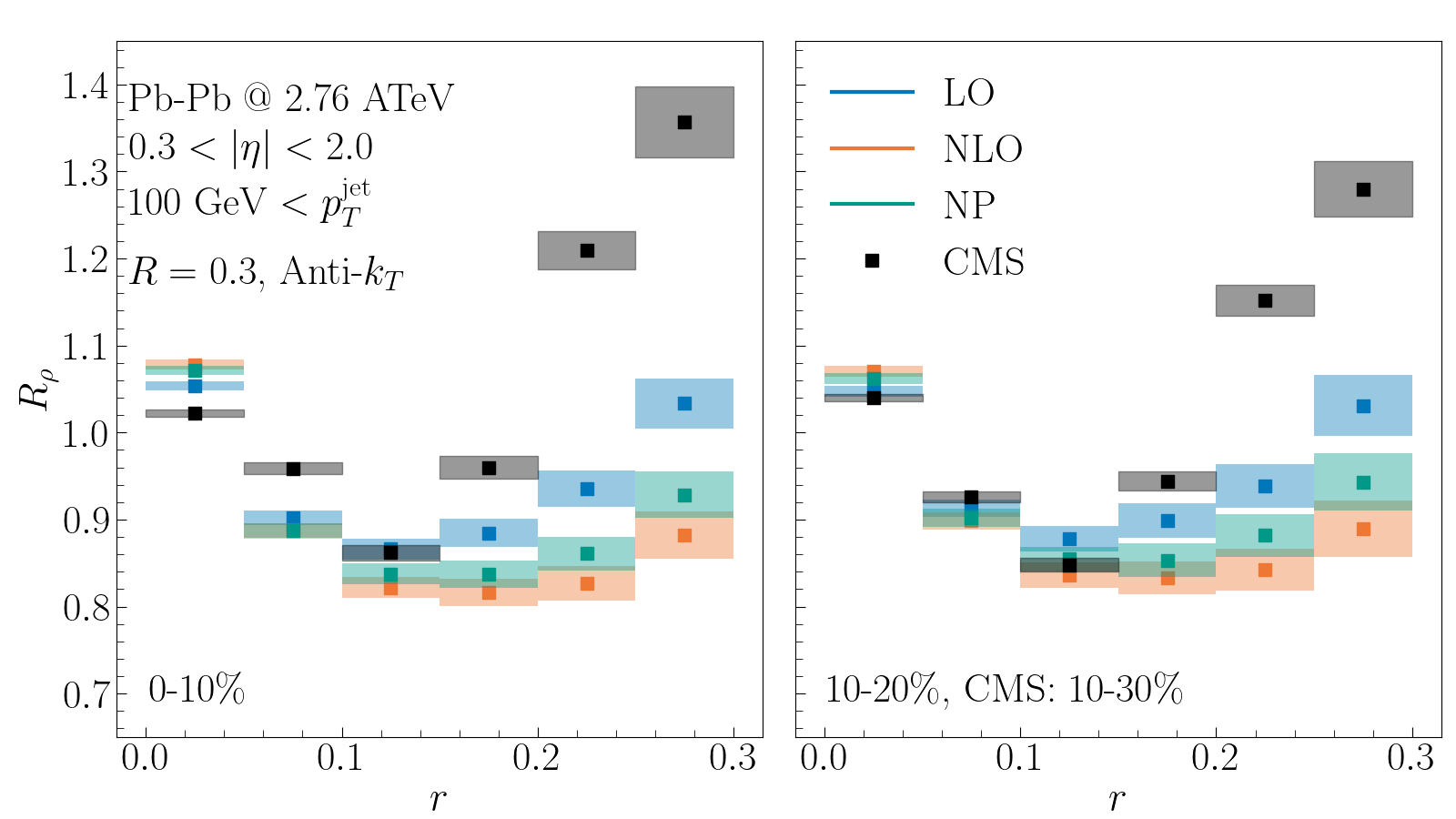}
        \caption{Jet shape ratio for two centrality classes of \pbpb\, collisions at $\sqrt{s}=2.76$ ATeV. Jets are clustered using the anti-$k_T$ algorithm for $R=0.3$ and the observable is calculated for jets in $0.3 < |\eta|<2.0 $ and $p^{\mathrm{jet}}_T> 100$ GeV. The data from the $10$-$30\%$ is compared to $10$-$20\%$ centrality simulation results. Data taken from the CMS Collaboration~\cite{CMS:2013lhm}.}
        \label{fig:coll.kernels.jet.shape.ratio}
    \end{figure}
    
    Fig.~\ref{fig:coll.kernels.jet.shape.ratio} shows the calculation of jet shape ratio, Eq.~\ref{eq:jet.shape.ratio}, for two centrality classes and compared to data from the CMS Collaboration. We see very good agreement between the three models when close to the jet axis ($r\leq 0.1$) and a separation between the models with higher order kernels and the LO one in the outer annuli. This observation is particularly acute for the $0$-$10\%$ centrality, while in $10$-$20\%$ class the theory calculations move a bit closer to each other. The (relatively) steep rise in the data toward the periphery of the jet, however, is not reproduced in any of the models. This is due to the setup of our current simulations and the influence of the development of the parton shower in the high-virtuality stage and the subsequent low-virtuality energy loss. See App.~\ref{sec:compare.jetscape.matter} for a short discussion on this and comparison to simulations with \matter. 
     
    We can see the effect of the rates in populating the lower energy modes or their propensity for energy loss via soft (relative to jet energy) radiation. In other words, the NLO and NP rates emit more (relatively) soft partons which get pushed out of jet cone while the LO does not do so. This is similar to the scenario with a fixed-\alphas, discussed in Ref.~\cite{Yazdi:2022bru}. There, when re-scaled by a constant factor, the NLO and NP rates could be made to match the LO rate for hard radiation though significant differences remained for soft radiation.  

    \begin{figure}
        \centering
        \includegraphics[width=\linewidth]{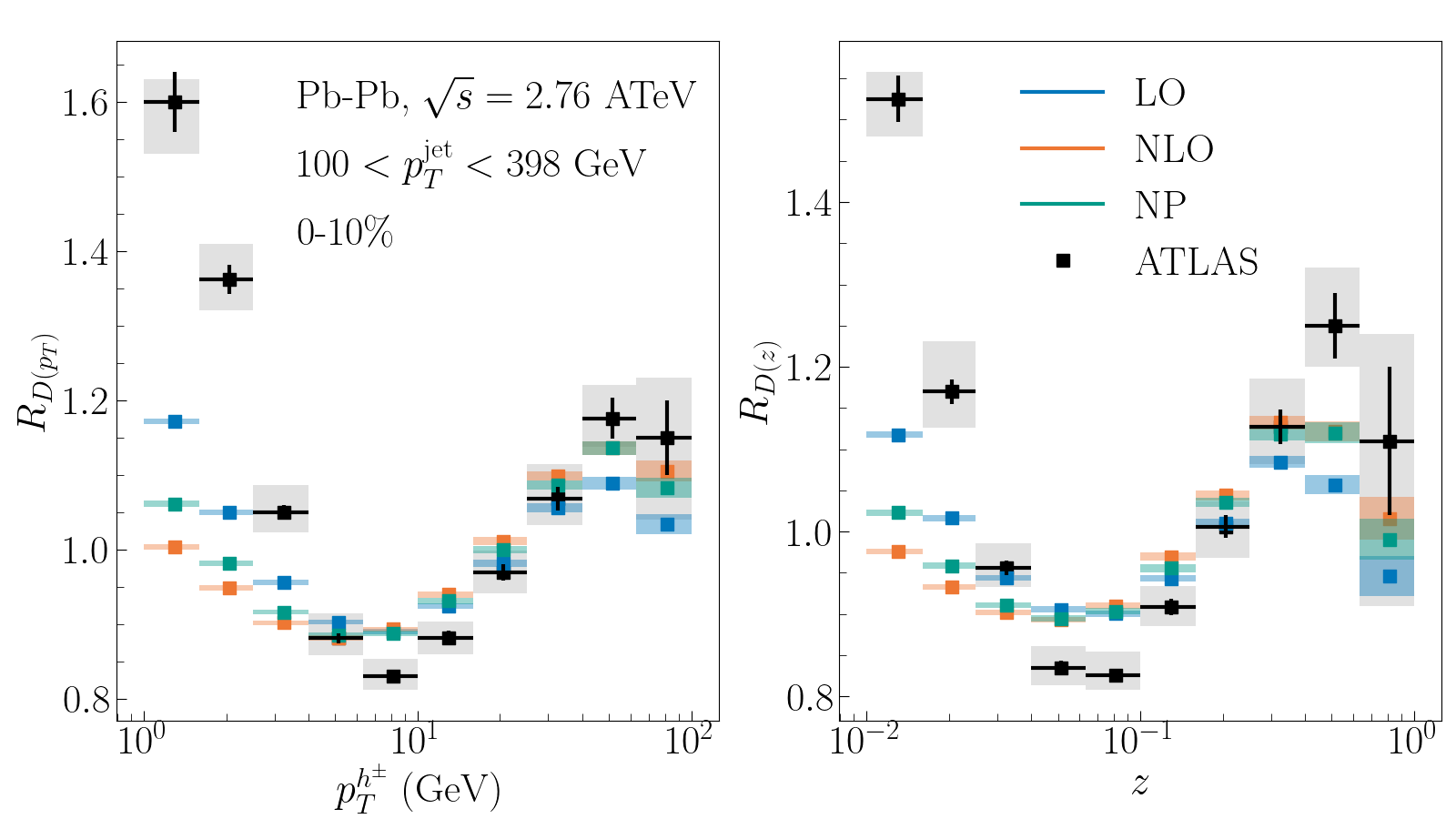}
        \caption{Jet fragmentation function ratio for charged hadron constituents for \pbpb\, collisions at $\sqrt{s}=2.76$ ATeV at $0$-$10\%$ centrality. Jets are clustered for $R=0.4$ using the anti-$k_T$ algorithm and those at midrapidity, $|\eta|<2.1$ with $100<p^{\mathrm{jet}}_T<300$ GeV are used in the calculation. Data from the ATLAS Collaboration~\cite{ATLAS:2017nre}.}
        \label{fig:coll.kernels.jet.FF.ratio}
    \end{figure}

    The final observable to consider is the jet fragmentation function ratios for charged hadrons presented in Fig.~\ref{fig:coll.kernels.jet.FF.ratio} and computed using Eqs.~\eqref{eq:FF.z} and~\eqref{eq:FF.pT}. The same qualitative features as the jet shape ratios of Fig.~\ref{fig:coll.kernels.jet.shape.ratio} are also noticeable here. The NLO and NP results are very close to each other, and visibly separate from the LO results. There is a flip in the order of the curves at around $z\approx 0.1$ or charged hadron $p_T\approx 10$ GeV. Below these cuts, LO results indicate a larger population of soft hadrons relative to the other two while it sees a slightly smaller number of the leading, hard hadrons. This again follows from the discussion of the propensity of the different rate sets in emitting soft radiation from the incoming hard parton and the ability of elastic channels in scattering these soft partons away from the leading parton.

%% file: conclusion.tex
\section{Conclusions \& Outlook}
Jet energy loss and the analysis of jet shapes remain important signals of the creation of the QGP. For light quarks and gluons, the dominant modes of energy loss are inelastic splittings and elastic scatterings with the particles of the medium. It is found that a detailed, multi-observable study of the new collision kernels in a dynamically evolving QGP requires particular attention to how the low-virtuality evolution in medium is seeded. As shown in Sec.~\ref{sec:fit.and.formtime}, in the absence of a formation time in the high virtuality shower, there are no combination of parameters controlling \alphas\, of the two energy loss channels that would admit a simultaneous description of charged hadron and jet \raa. The situation improves immediately, when formation time is introduced to the high virtuality parton shower. One can achieve a simultaneous description of the charged hadron \raa\, and jet \raa. Further more, the performance of the model against jet shape ratio and jet fragmentation-function ratios is also improved. In particular, this improvement is observed close to the jet axis for the former and for the leading charged hadrons for the latter. The results of Sec.~\ref{sec:fit.and.formtime} show that the significance of shower formation time in the high virtuality stage of a \martini\, calculation is parameter-independent. Given the similar observation in Ref.~\cite{he:2022evt} for the \lbt\, model, we can state that necessity of the inclusion of a shower formation time is also \textit{model}-independent. 

While shower formation time allows for a simultaneous description of charged hadron and jet \raa, we can also see the limitation of using the present prescription for it in jet substructure observables. The model does not provide a good description for the softer segments of the substructure observables, $r\geq 0.1$ in jet shape ratio or $z\geq 0.1$ in jet fragmentation function ratios. As the view of formation time in this study is still somewhat schematic, it underestimates the amount of interactions experienced by evolving parton distribution. In a more realistic scenario, the radiating system can experience interactions with the medium, potentially receiving strong enough ``kicks'' from the medium to significantly reduce the formation time of the radiated parton. This would then percolate along the shower tree. Furthermore, also allowing for energy loss in the process would alter the evolving spectrum which would then pass to the low-virtuality stage of evolution. The next generation of heavy-ion event generators should address these different aspects in a scenario where the fluid dynamical and the shower evolution proceed holistically.
Another point to consider in this discussion is the modification of the running of the strong coupling, with and without the inclusion of formation time in the high virtuality parton shower. As mentioned in Sec.~\ref{sec:fit.proc.and.res}, and clearly demonstrated in Fig.~\ref{fig:compare.alphas.LO}, the running coupling is significantly larger when evaluated for a time-delayed parton shower. Given the importance of \alphas\, as not just a free parameter but also a gauge of the perturbative-ness of the system, the issue of the faithful modelling of the high virtuality stage -- a dynamic formation time and energy loss -- is crucial in the better understanding of the jet-medium coupling. A similar observation and conclusion was also reached in Ref.~\cite{Ke:2020clc} for a different transport model. 

In the second part of this study, we analyzed the phenomenological influence of the higher order collision kernels on the energy loss of hard partons, particularly in comparison to the LO kernel using the AMY framework. This in turn required the optimization of the parameters of \martini\, that govern the scale dependence of the running coupling. After achieving a simultaneous description of the nuclear modification factors of 
charged hadrons and jets, we performed comprehensive parameter scans 
of \martini\,using leading-order (LO), up-to-next-to-leading-order (NLO), and 
non-perturbative (NP) collision kernels. Taking the optimized parameter sets 
respectively, we observe that three kernels exhibit remarkable similarities 
--- in terms of the overall value along with transverse-momentum and centrality dependence --- 
in charged hadron \raa\, and jet \raa. In contrast to modifications to the overall 
jets or hadrons, we observe sizable differences in the modification of jet 
substructure observables, i.e., jet radius distribution and fragmentation function. 
Such differences are caused by the difference in the radiation rates of relatively 
soft gluons what survive in the evolution in medium. Given the sensitivity of the 
results to the initial condition of the evolving parton distribution, the differences 
require further study where the high-virtuality stage is modelled in a more dynamic way. 

Since the differences between rates from different collision kernels are more 
readily observed in the low to intermediate range of transverse momentum~(See App.~\ref{sec:pythia.guns}), 
it would be useful to consider probes within that region of $p_T$. We anticipate that mini-jets~\cite{Pablos:2022piv} 
and electromagnetic probes would be good candidates. In particular, the jet-medium photon 
spectrum~\cite{ModarresiYazdi:2023cby} has been shown to be a sizeable portion of the total photon spectrum in this $p_T$ region and has the benefit of being (nearly) directly proportional to the evolving quark and 
anti-quark spectrum while not experiencing the effects of hadronization. This is also a topic for future work.

%% file: appendix_comp_matter.tex
\section{Initial state with MATTER in a JETSCAPE framework}\label{sec:compare.jetscape.matter}
    \matter\, is a state-of-the art model of the initial high virtuality stage of the evolution of hard partons. Here we present a comparison of the LO rate sets in \martini\, as a standalone framework for energy loss, with and without a time delayed parton shower and \martini\, as a model of low-virtuality energy loss and a module in a \jetscape\, framework. The former setup is what has been studied in Sec.~\ref{sec:fit.and.formtime}. The case of \martini\, as a component of a multi-stage model is handled via \jetscape, where the hard event and initial state parton shower, including cold nuclear effects, are simulated by \pythia. The high virtuality hard partons are then evolved by \matter\, which modifies the DGLAP splitting kernel by including medium effects,
    \begin{equation}\label{eq:matter.dglap.simple}
        P^a(y, Q^2) = P^{a}_{\mathrm{vac.}}(y) + P^{a}_{\mathrm{med.}}(y, Q^2)
    \end{equation}
    where $Q^2$ is the virtuality of the incoming parton of species $a \in (q,\bar{q},g)$ and light-cone momentum $p^+$ which splits to two partons carrying $yp^+$ and $(1-y)p^+$ of the forward light-cone momentum. In the absence of a medium, the second term in Eq.~\ref{eq:matter.dglap.simple} is zero and a \matter\, evolved final state parton shower becomes identical to one generated by \pythia. Otherwise, energy loss in \matter\, is controlled by the HTL expression for $\hat{q}$, as calculated in Ref.~\cite{He:2015pra}
    \begin{align}
            \hat{q} = C_{A}\frac{42\;\zeta(3)}{\pi}\alpha^2_s T^3&\bigg[\ln{\left(\frac{5.7 E T}{4 m^2_D}\right)}\Theta(E- 2 \pi T)+\nonumber\\
            &\ln{\left(\frac{5.7\left(2\pi T^2\right)}{4 m^2_D}\right)}\Theta(2\pi T-E)\bigg]
        \label{eq:qhat.expression.MATTER}
    \end{align}
    where $\zeta(3)$ is the Ap\'ery constant ($\zeta(3) = 1.20205$) and $\Theta$ is the Heaviside function. 

    High virtuality partons are then evolved down to a virtuality cut off, $Q_{min} = 2$ GeV. Partons with virtuality above this cutoff are considered high-virtuality and those below are taken to be on the mass shell. On shell partons are then handed over to \martini\, for further evolution if they are still within the medium. In the results that will be shown in this section, we fix the momentum cut off for energy loss, recoil and radiative events to $p_{\mathrm{cut}}=2$ GeV. In testing, we found that this value is nearly equivalent to a $p_{\mathrm{cut}}=4T$ cut that was used in the rest of this work. 

    We note that the underlying hydrodynamic simulations of the soft sector in what we compare here are different. For the \jetscape\, framework results presented here, the QGP evolution is modelled by a \trento+free-streaming+\vishnu(2+1)D evolution, where the various parameters are fixed in a landmark Bayesian study of heavy ion collisions~\cite{Bernhard:2019bmu}. The discussion of the differences between this framework and the \ipg+\music(3+1)D view of the QGP evolution, which we used in the body of the text, is beyond the scope of this work. We only note that both successfully reproduce a great number of soft sector observables and that the jet evolution and energy loss, at least for the leading hadrons, is not sensitive to the differences in the modelling of the soft sector.
    
    To study the effect of incorporating spacetime information into the high virtuality parton shower shower stage of evolution, we compare the \martini\, results of Sec.~\ref{sec:fit.and.formtime}, with and without a formation time in the parton shower, to results from a \jetscape\, calculation, where \matter\, handles the final state shower. In total, we consider three different setups for the multi-stage simulations where \matter\, handles the high virtuality stage
        \begin{enumerate}
            \item \matter\, ($\hat{q}\neq 0$) + \martini\, where $\kappa_0\equiv(\kappa_r,\kappa_e)=(1.5, 4.5)$
            \item \matter\, ($\hat{q}\neq 0$) + \martini\, where $\kappa_1\equiv(\kappa_r,\kappa_e)=(2.0, 8.6)$
            \item \matter\, ($\hat{q}= 0$, or ``vac'')+ \martini\, with $\kappa_1$.
        \end{enumerate}
    In the above, $\kappa_0$ parameter set for the running coupling in \martini\, had been previously~\cite{Park:2019sdn, Shi:2022rja} found to provide good agreement with the data when used in a multi-stage \jetscape\, framework. We compare this parameter set to the ``optimal'' parameter set that was found for \martini\, simulations without formation time, $\kappa_1=(2.0, 8.6)$. This latter parameter was also used in calculations where the \matter\, phase did not allow for energy loss, i.e. $\hat{q}=0$. 

    \begin{figure}
        \centering
        \includegraphics[width=\linewidth]{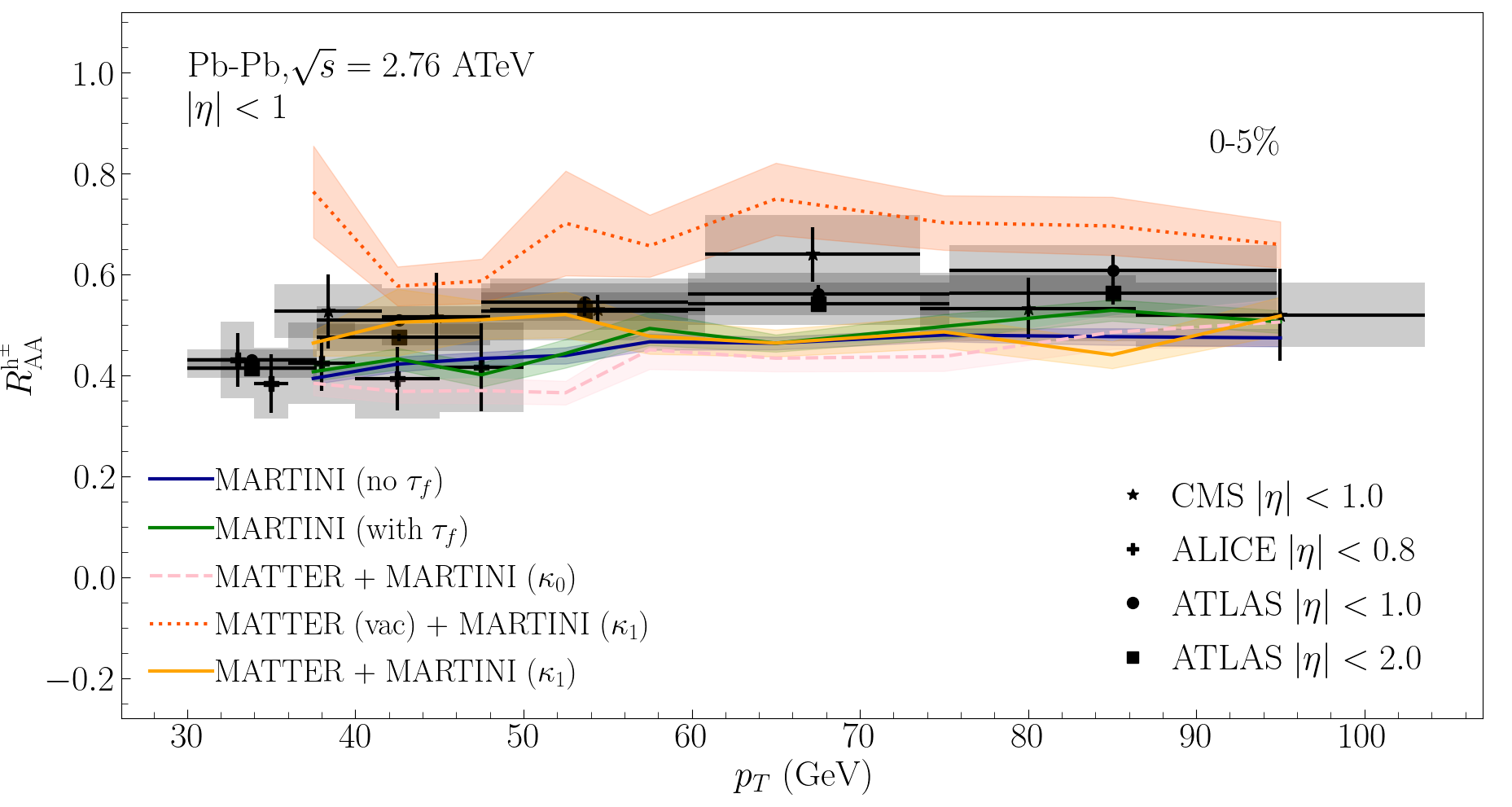}
        \caption{Comparison of the effect of high virtuality energy loss on charged hadron \raa, at $0$-$5\%$ centrality class of \pbpb\, collisions with $\sqrt{s}=2.76$ ATeV. Lines for \martini\, with and without formation time are the same as previously shown in Sec.~\ref{sec:fit.and.formtime}. All other results are computed using the \jetscape\, framework, where \martini\, accounts for one stage of evolution. Data CMS~\cite{CMS:2012aa}, ATLAS~\cite{ATLAS:2015qmb} and ALICE~\cite{ALICE:2012aqc} Collaborations.}
        \label{fig:charged.raa.martini.vs.matter}
    \end{figure}

    Fig.~\ref{fig:charged.raa.martini.vs.matter} shows the result of calculations of \martini, with and without shower formation time, along side those which include \matter\, in their evolution. As before, we see that the charged hadron \raa\, is quite insensitive to the details of the simulation and that all four of the five curves easily reproduce the data. The \matter+\martini\, results that include energy loss in their \matter\, phase, regardless of which $\kappa$-set is used in their \martini\, stage ($\kappa_0$ vs. $\kappa_1$) fall neatly on top of the \martini-only results. The case where \matter\, is in vacuum mode with $\hat{q}=0$ is predictably above all other curves, since there is no energy loss in its high-virtuality stage and the running $\alpha_{s}$ in its \martini\, stage is weaker relative to the \matter($\hat{q}\neq 0$)+\martini($\kappa_1$) simulations. 

    \begin{figure}
        \centering
        \includegraphics[width=\linewidth]{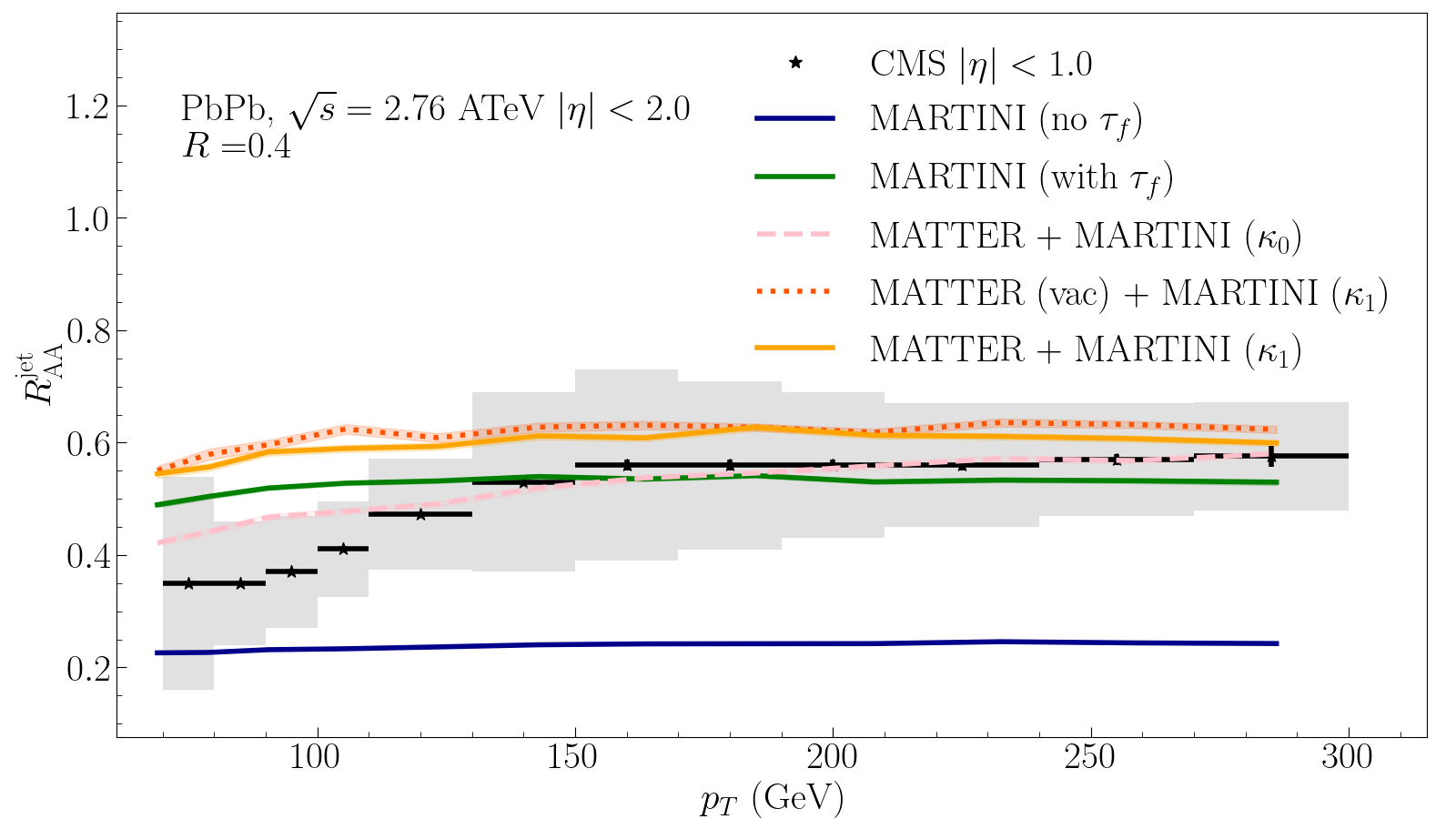}
        \caption{Similar to Fig.~\ref{fig:charged.raa.martini.vs.matter}. Comparison of the effect of high virtuality energy loss on inclusive jet \raa, at $0$-$5\%$ centrality class of \pbpb\, collisions with $\sqrt{s}=2.76$ ATeV. Data from the CMS Collaboration~\cite{CMS:2016uxf}.}
        \label{fig:jet.raa.martini.vs.matter}
    \end{figure}

    Using the same events that resulted in the charged hadron results of Fig.~\ref{fig:charged.raa.martini.vs.matter}, we can construct the inclusive jet \raa, given in Fig.~\ref{fig:jet.raa.martini.vs.matter}, for jet cone radius of $R=0.4$. \martini-only simulations where the parton shower is assumed to have developed before entering the thermal medium, denoted as \martini\, without formation time, is the visible outlier. Including formation time, using the prescription described in Sec.~\ref{sec:fit.and.formtime}, places the \martini\, simulation among the other three simulations that have \matter\, in their high virtuality state. 

    \begin{figure}[!ht]
        \centering
        \includegraphics[width=\linewidth]{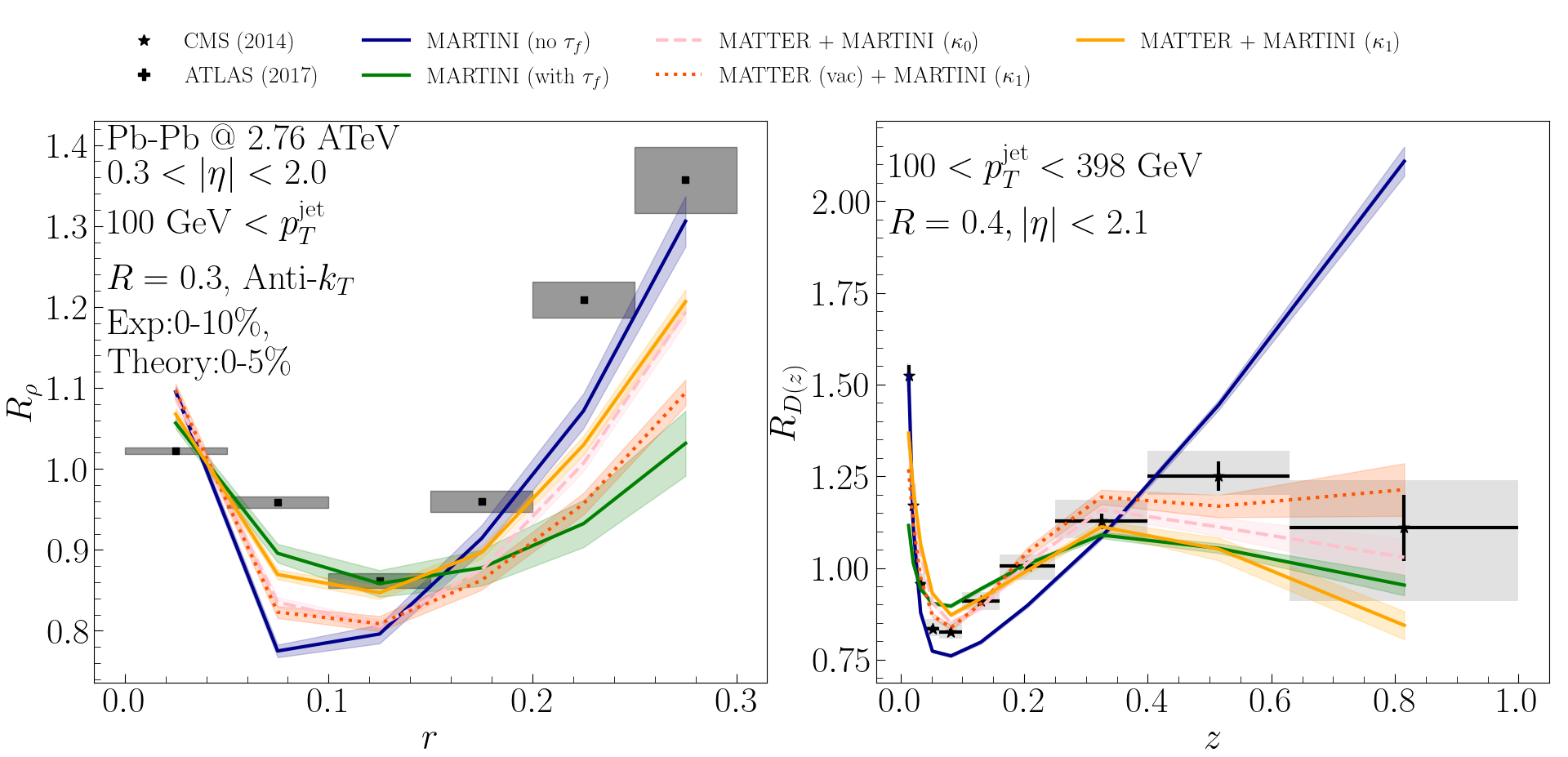}
        \caption{Similar to Figs.~\ref{fig:charged.raa.martini.vs.matter} and~\ref{fig:jet.raa.martini.vs.matter}. Comparison of the effect of high virtuality energy loss on jet substructure for \pbpb\, collisions with $\sqrt{s}=2.76$ ATeV and $0$-$5\%$ centrality. On the left, jet shape ratio for high-$p_T$ jets with $R=0.3$ and $0.3<|\eta|<2.0$ is presented while the figure on the right shows the jet fragmentation function ratios for charged hadrons as a function of their momentum fraction. Data from the CMS~\cite{CMS:2013lhm} and ATLAS~\cite{ATLAS:2017nre} Collaborations, for jet shape ratio and jet fragmentation function ratio, respectively.}
        \label{fig:jet.substruct.martini.vs.matter}
    \end{figure}
    
    Finally, jet substructure observables as presented in Fig.~\ref{fig:jet.substruct.martini.vs.matter}, where we consider jet shape ratio (right) and jet fragmentation function ratio (left). The models which include \matter\, as a component of their evolution are sandwiched between the \martini-only models. Close to the jet axis, models with some notion of spacetime are much closer to the data, and more \pp-like. As we move away from the jet axis, we deviate sharply from the vacuum jet shape. The model which best captures this situation is \martini, without formation time. Though \matter+\martini\, models which include energy loss in their \matter\, phase also exhibit a similar rapid rise. 
    The jet fragmentation function ratio as a function of the charged hadron momentum fraction. As shown previously in Sec.~\ref{sec:runs.with.fitted.params}, \martini\, without formation time can match the small $z$ ($z<0.1$) part of this observable and even seems to have a good description of the slope and rise of the data for $0.1 < z < 0.4$. However, for the leading or more energetic hadrons with $z \geq 0.5$, the FF ratio from a \martini\, with long evolution time exhibits a rising ratio and significant modification relative to a \pp\, baseline while all other models which include spacetime information, with or without energy loss in their high virtuality stage, cut off this rise in $R_{D(z)}$. 
    

%% file: appendix_coll_kern_long_etime.tex
\section{Higher order kernels and formation time of shower}\label{sec:long.short.formtime.kernels}

    We discussed the observed distances between the different rate sets that arise from using LO, NLO and NP collision kernels in Sec.~\ref{sec:higher.order.kernels}. There, we only focused on the simulations with the simple prescription of formation time in their parton shower. Here, we provide a comparison of those calculations to the simulation results where there is no formation time on the parton shower. This is done only for the jet substructure observables, as the charged hadron and jet \raa\, calculations are not particularly instructive and simply reinforce our conclusions from Figs.~\ref{fig:compare.charged.raa.formation.time.final} and~\ref{fig:compare.jet.raa.formation.time.final}. 

    \begin{figure}
        \centering
        \includegraphics[width=\linewidth]{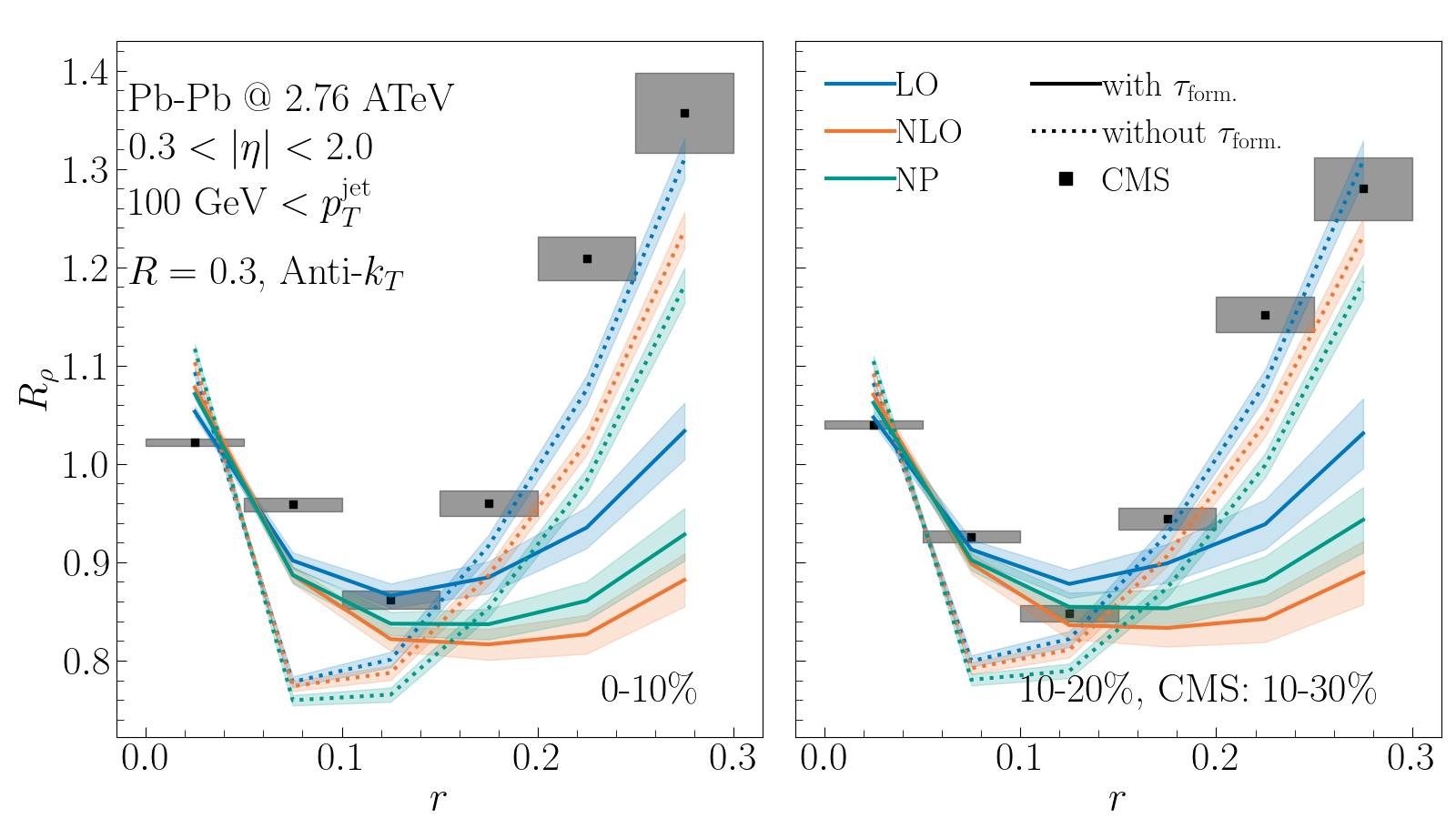}
        \caption{Jet shape ratio comparing with and without formation time. Results are compared to data from the CMS~\cite{CMS:2013lhm} Collaboration.}
        \label{fig:all.kernel.comp.formation.time.appendix.jet.shape}
    \end{figure}

    We previously saw in Sec.~\ref{sec:runs.with.fitted.params}, when considering only the LO rate set, the model with formation time does a much better job of matching the data close to the jet axis, while the outer annuli are better captured by the model without formation time. There, we also noted the caveat that the model without formation time is operating with the \textit{wrong} jet population and should be taken as an un-realistic but illustrative \textit{long evolution time} limit of jet energy loss. Here, in Fig.~\ref{fig:all.kernel.comp.formation.time.appendix.jet.shape}, we see that the change in the collision kernel, from LO to NLO or NP, does not change this observation. Without formation time, the shape ratios begin to separate at $r\approx 0.08$, with an ordering where $R^{\mathrm{LO}}_{\rho}>R^{\mathrm{NLO}}_{\rho}>R^{\mathrm{NP}}_{\rho}$. In the case with formation time, the separation occurs at the same value of $r$, but the ordering is different in this case: while the LO results are still above the other two, the NLO result is now at the bottom and the NP curve is situated between the LO and NP curves. 
    
    \begin{figure}
        \centering
        \includegraphics[width=\linewidth]{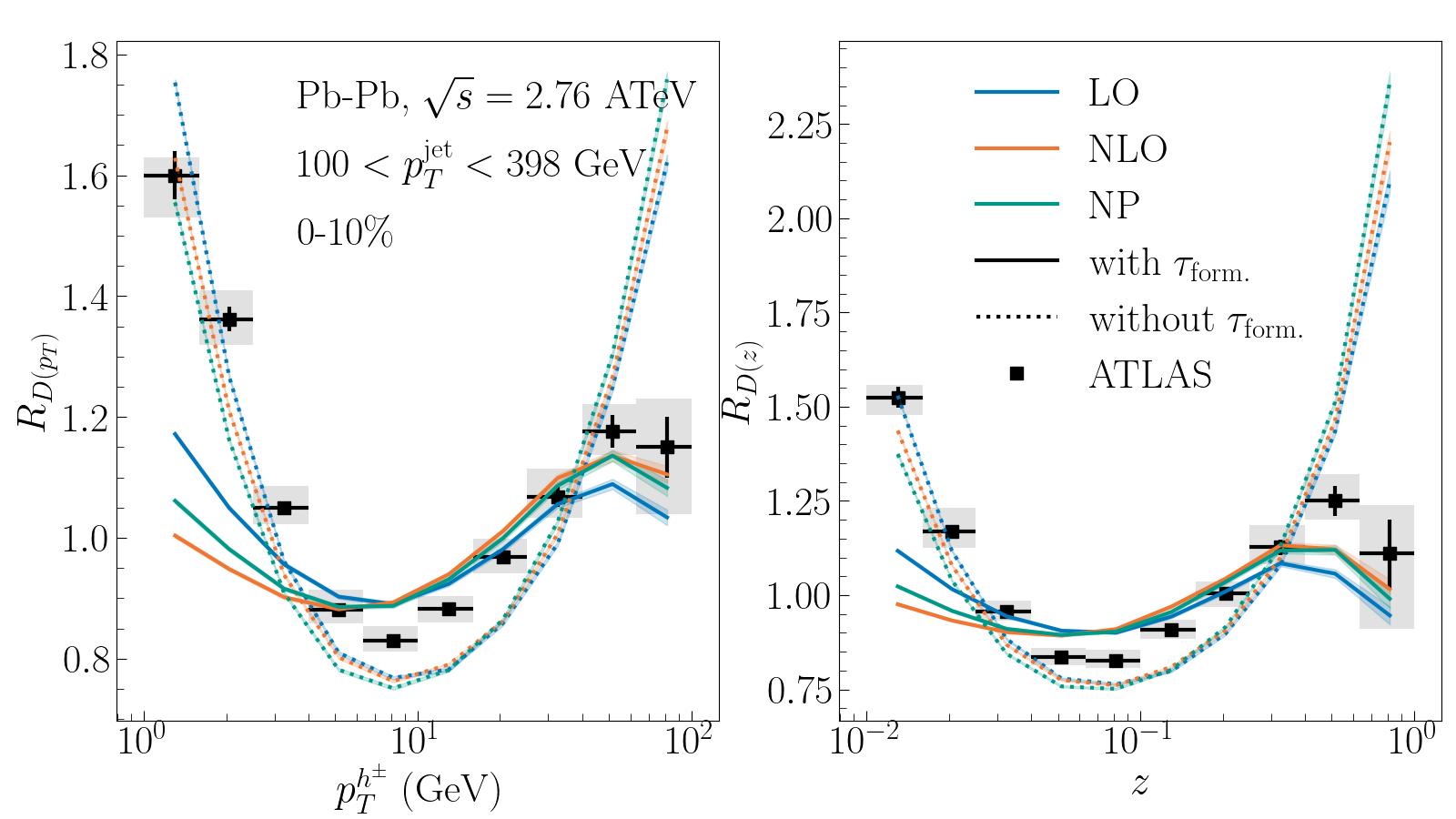}
        \caption{Jet fragmentation function ratio comparing with and without formation time. Results are compared to data from the ATLAS~\cite{ATLAS:2017nre} Collaboration.}
        \label{fig:all.kernel.comp.formation.time.appendix.jet.FF}
    \end{figure}

    A similar effect is observed in Fig.~\ref{fig:all.kernel.comp.formation.time.appendix.jet.FF}, for jet fragmentation function ratios. When shower formation time is included, simulations with NLO and NP kernels are very close to each other and separate from the LO results. Comparing these curves with their corresponding instantaneous shower, we again see the flipping of the relative order. For the leading charged hadrons, in calculations without a formation time we can see that $R_{\mathrm{NP}}>R_{\mathrm{NLO}}>R_{\mathrm{LO}}$ while when formation time is included, $R_{\mathrm{NLO}}>R_{\mathrm{NP}}>R_{\mathrm{LO}}$. Similar to the jet shape ratio, here we again see improvement in capturing the quenching of the leading hadrons while the agreement is not as good for softer modes with $p_T \leq 20$ GeV. The ultra soft modes with $p_T\leq 3$ GeV are much better captured by the simulations without formation time while being missed completely by the simulations with formation time. As with jet shape ratio, this is indicative of the importance of energy loss in the high-virtuality stage of the shower.

%% file: appendix_pythia_guns.tex
\section{Single Shower Examples}\label{sec:pythia.guns}
    \begin{figure}
        \centering
        \includegraphics[width=\linewidth]{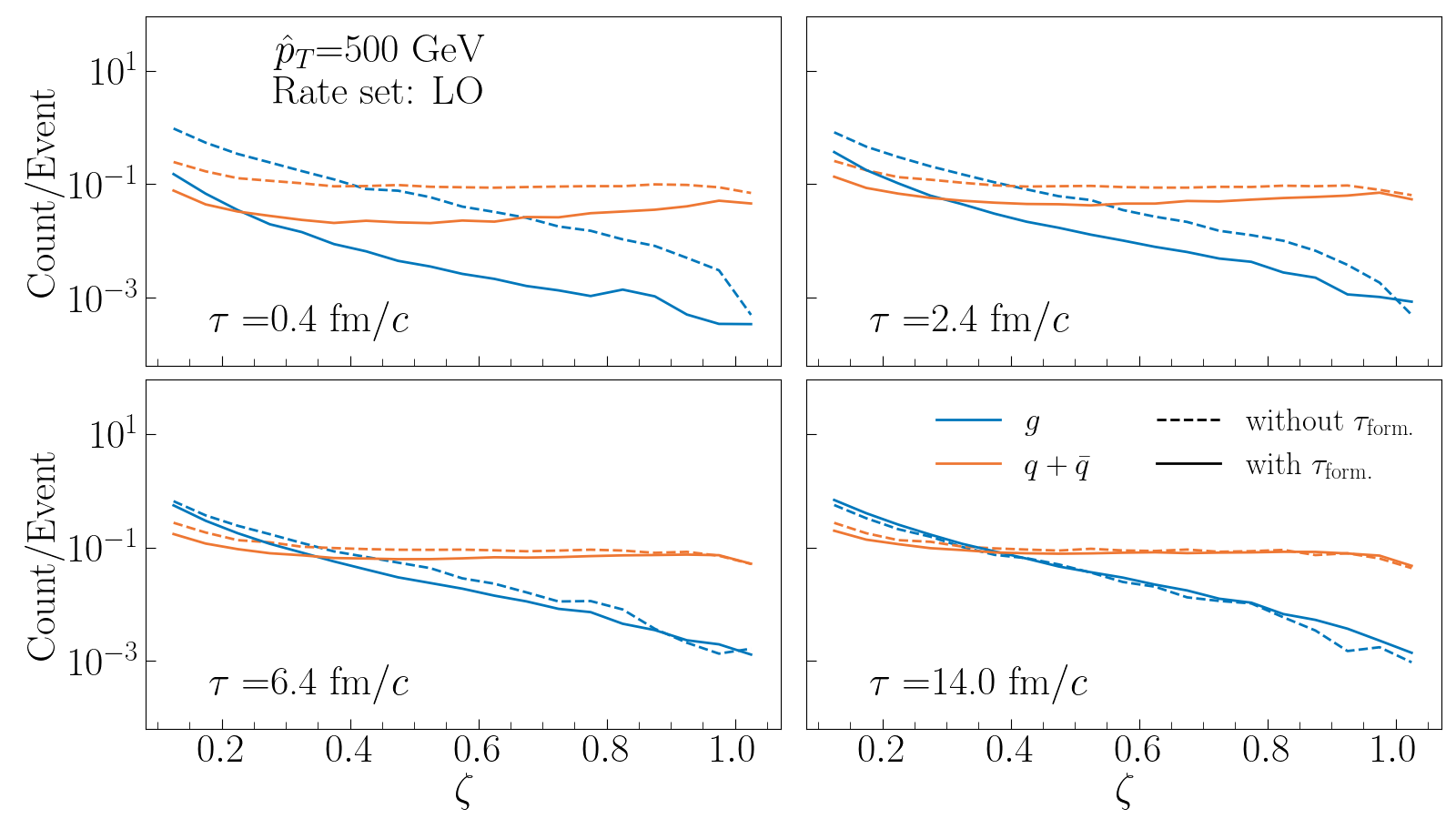}
        \caption{A \martini\, event generated for $\hat{p}_T = 500$ GeV using the LO rate set. Partons are binned in the pseudorapidity window $|\eta|<2.0$. See text for details.}
        \label{fig:gun.500GeV.LO.formtime.study}
    \end{figure}
    
The results shown in this work all concerned strongly interacting probes, after they had already been hadronized. We can also consider the evolution of a parton shower from the hard interaction point, through the QGP. This would be similar to a QGP brick study, though in this case, we are more interested in the evolution of a complete shower through a dynamically evolving QGP. The benefit of this type of study are twofold. First, one can clearly observe effect of formation time of the high virtuality shower on the evolution of the parton spectrum. Second, we analyze the differences between a parton shower evolution via rates computed using different collision kernels in a more controlled way while maintaining an element of realism. 
    \begin{figure}
        \centering
        \includegraphics[width=\linewidth]{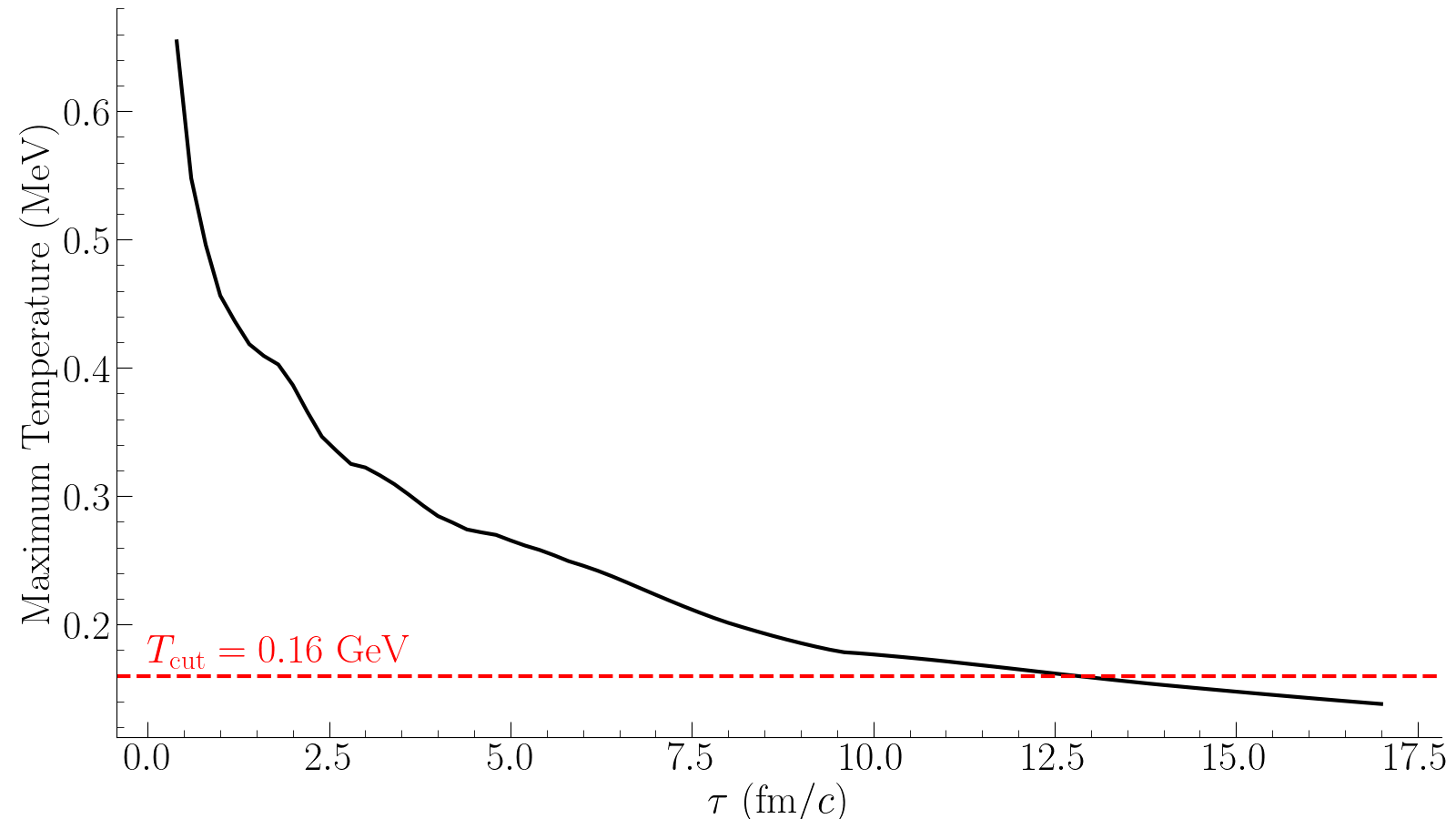}
        \caption{Evolution of maximum temperature in the most central collision of \pbpb\, at $2.76$ A TeV, at midrapidity ($\eta=0$). The hard temperature cutoff used in jet energy loss calculations, $T_{\mathrm{cut}}=160$ MeV, is shown by a horizontal line. See text for details.}
        \label{fig:temp.evol.central.event}
    \end{figure}
We generate \pythia\, events for a given value of invariant transverse momentum at the partonic level of the hard scattering, $\hat{p}_T$, with and without shower formation time. The generated parton showers are then evolved in the QGP medium, using the optimized parameters for \alphas, with all other momentum and temperature cut-offs held exactly the same as the simulations presented in the main body of this work. For a selection of time steps in the evolution of the QGP, the gluon and $q+\bar{q}$ distributions are binned in histograms. These histograms are constructed for the variable $\zeta$, defined as
    \begin{equation}
        \zeta_i \equiv \frac{p_{T,i}}{\hat{p}_T}
        \label{eq:zeta.def}
    \end{equation}
where $p_{T,i}$ is the transverse momentum of a given parton in the developing parton shower. The pseudorapidity window in which the partons are collected is for $|\eta|<2.0$, corresponding (approximately) to the pseudorapidity window used for charged hadron and jet \raa\, results that were presented previously.  
        
        \begin{figure}
            \centering
            \includegraphics[width=\linewidth]{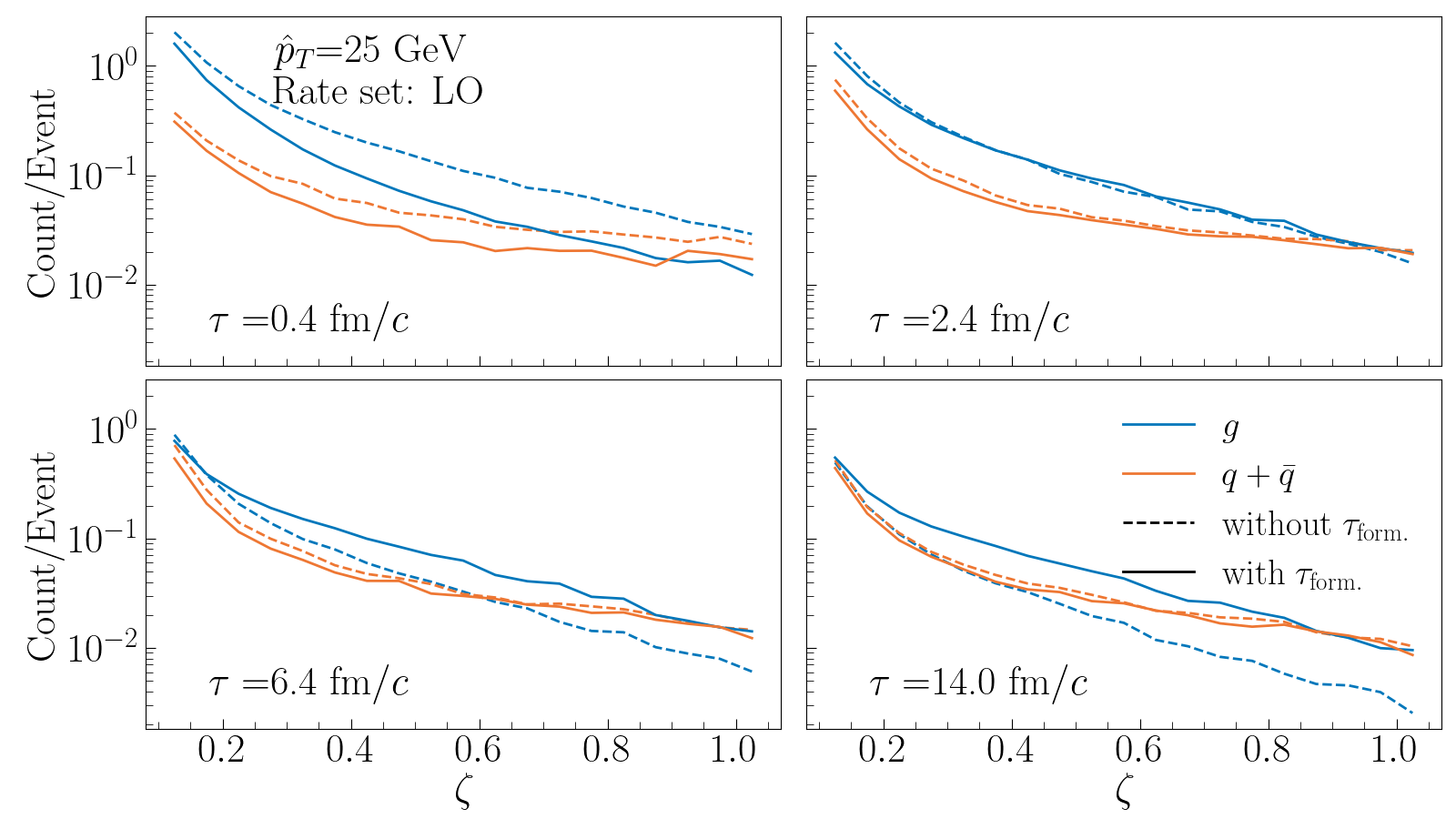}
            \caption{Same as configuration as fig.~\ref{fig:gun.500GeV.LO.formtime.study}, for $\hat{p}_T = 25$ GeV.}
            \label{fig:gun.25GeV.LO.formtime.study}
        \end{figure}
        
Fig.~\ref{fig:gun.500GeV.LO.formtime.study} shows the evolution history of an event with $\hat{p}_T=500$ GeV. As expected, the initial distribution of hard partons that enter the plasma are quite different in the two models. This is due to the implementation of formation time in the final state radiation of the high virtuality shower. For each sub-figure, the parton distributions with formation time, include only partons whose $\tau_{\mathrm{form.}} \leq \tau$, where $\tau$ is the current hydro time. The final subplot at $\tau=14$ fm$/c$ can be thought of as the final distribution exiting the plasma towards the hadronization stage. This is because in our calculations, we take the decoupling temperature of the jet from the hydrodynamic background to be $160$ MeV, as previously stated. Looking at the evolution of the maximum temperature at midrapidity, Fig.~\ref{fig:temp.evol.central.event}, it is clear that by $\tau=14$ fm$/c$, the hard partons are frozen out of evolution as the maximum temperature in the plasma is below the temperature cutoff. The gluon and quark distributions for simulations with and without formation time, start very different from each other but by the end of the evolution, the high-$p_T$ region of the distributions are nearly identical, with only small differences observable in the lower $\zeta$ region. This is in agreement with the results shown in the study of jet shape and jet fragmentation function ratios in Sec.~\ref{sec:fit.and.formtime}. However, in those observables, the differences were more pronounced for much softer partons. Here, $\zeta\leq0.2$ corresponds to partons of $p_T\leq 100$ GeV, which are still quite hard. Thus we can turn our focus to a shower structure from a much softer collision event, $\hat{p}_T=25$ GeV. This is shown in Fig.~\ref{fig:gun.25GeV.LO.formtime.study}. Here, we begin to see an interesting evolution history for the evolving hard parton spectra. By $\tau=2.4$ fm$/c$, the models with and without formation time have converged to each other, having started from very distinct starting positions. They diverge again, and by the end of the evolution, it is clear that the difference between a model with formation time in the high virtuality shower and one without, is in the gluon population. Furthermore, the gluon and quark populations in calculations without formation time arrive at similar distribution of partons for $\zeta\leq0.3$ or $p_T\leq 7.5$ GeV. 

    \begin{figure}
        \centering
        \includegraphics[width=\linewidth]{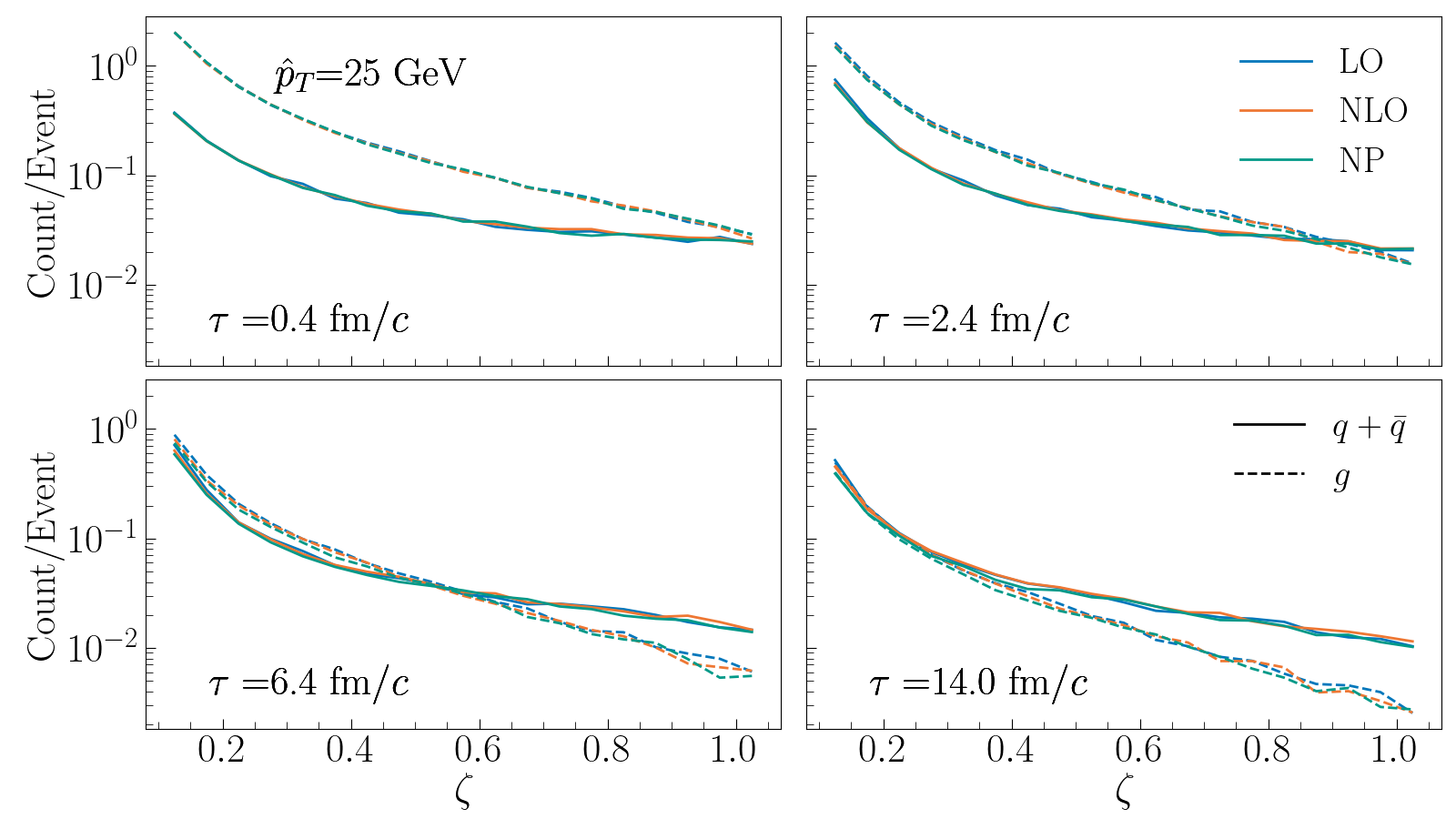}
        \caption{Results for the evolution of a $\hat{p}_T=25$ GeV shower, without formation time in the high virtuality stage for the three rate sets studied in this work. The same kinematic cuts as Fig.~\ref{fig:gun.500GeV.LO.formtime.study} are applied here.}
        \label{fig:guns.rate.comparison.25GeV.without}
    \end{figure}
    
Finally, we consider the different rate sets with these \pythia\, guns. Fig.~\ref{fig:guns.rate.comparison.25GeV.without} shows the evolution history of the $\hat{p}_T = 25$ GeV shower structure, without formation time in the high virtuality stage, for the three rate sets. By the end of the evolution, there resulting partonic spectra are nearly identical. Interestingly, the same convergence of gluon and quark spectra for soft partons that was observed in Fig.~\ref{fig:gun.25GeV.LO.formtime.study} is also visible here, for all three rate sets.    

        \begin{figure}
            \centering
            \includegraphics[width=\linewidth]{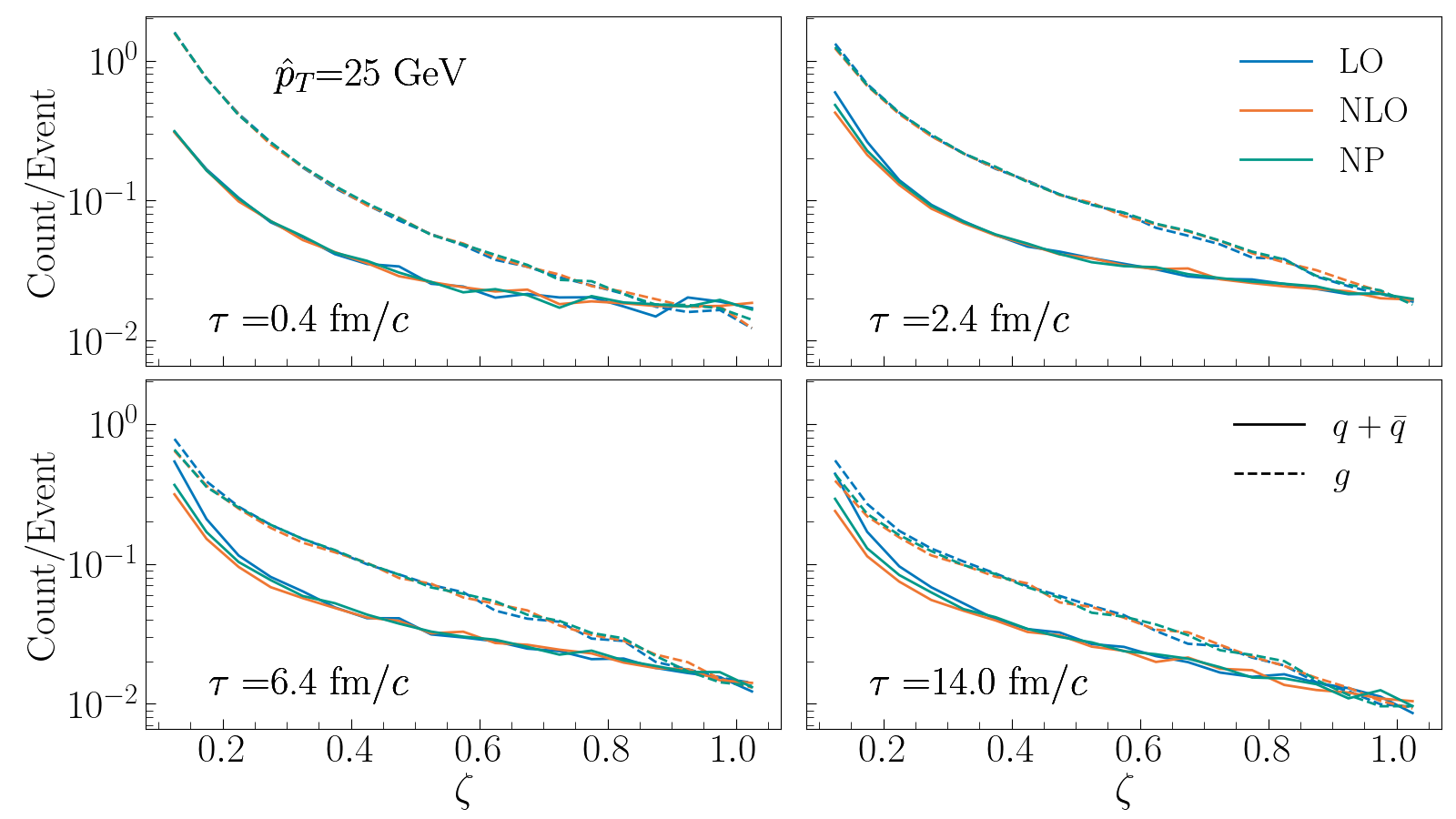}
            \caption{Same as Fig.~\ref{fig:guns.rate.comparison.25GeV.without}, but with formation time in the high virtuality stage.}
            \label{fig:guns.rate.comparison.25GeV.with}
        \end{figure}

Enforcing formation time in Fig.~\ref{fig:guns.rate.comparison.25GeV.with}, separates the gluon and quark distributions. The different rate sets still provide mostly similar results, with the differences most visible in the region of $\zeta\lesssim 0.4$, or $p_T \lesssim 10$ GeV. This difference is most pronounced in the quark spectrum (quarks + anti-quarks), and less so in the gluon spectrum. The relative difference in the overall size of the two spectra means the net difference between the three rate sets would be rather small for the hadronic observables. This latter step can also complicate the matters, as the non-perturbative process of hadronization which requires model building and fits to data, can further complicate the picture. In the body of this work, we showed the differences were significant and observable in the distribution of very soft hadrons ($p^{\mathrm{h}^{\pm}}_T < 10$ GeV) inside jets, having survived the hadronization stage, using a string hadronization model based on that of \pythia. 

This is where a complimentary probe which would be directly proportional and sensitive to the evolving parton spectrum would be useful. It has been shown that photons resulting from passage of quarks and anti-quarks through a QGP medium contribute significantly to the total photon yield, particularly in the $p_T$ region identified here ($p_T\lesssim10$ GeV)~\cite{ModarresiYazdi:2023cby}. Thus, one can use a multi-probe study, where the photon spectrum could provide (nearly direct) access to evolving $q+\bar{q}$ distribution, where one could observe differences between simulations using different collision kernels in the rates. 